\documentclass[a4paper,11pt]{article}

\usepackage[english]{babel}
\usepackage[T1]{fontenc}
\usepackage[utf8]{inputenc}
\usepackage{amssymb,amsmath,amsthm}
\usepackage{mathrsfs}
\usepackage{times}

\usepackage{amscd}
\usepackage{a4wide}
\usepackage{tikz}
\usepackage{physics}

\date{}

\usepackage[colorlinks=true,linkcolor= blue, citecolor=blue]{hyperref}

\numberwithin{equation}{section}
\newtheorem{theo}{Theorem}[section]
\newtheorem{cor}{Corollary}[section]
\newtheorem{ass}{Assumption}[section]
\newtheorem{prop}{Proposition}[section]

\newtheorem{defi}{Definition}[section]
\newtheorem{rem}{Remark}[section]
\numberwithin{figure}{section}

\newcommand{\ch}{H}
\newcommand{\cho}{H_0}

\newcommand{\br}{{\mathbb{R}}}
\newcommand{\bn}{{\mathbb{N}}}
\newcommand{\bc}{{\mathbb{C}}}
\newcommand{\spp}{{\mathcal{\textbf{\textup{S}}}_p}}
\newcommand{\sinf}{{\mathcal{\textbf{\textup{S}}}_\infty}}

\newenvironment{prof}
	{{\it Proof.}}
	{\hfill $\square$\vskip 8pt}

\title{Spectral non-self-adjoint analysis of complex Dirac, Pauli and Schr\"odinger operators of full rank with constant magnetic fields}
\author{Diomba Sambou$\footnote{Supported by the Chilean Fondecyt Grant 3170411.}$}

\begin{document}

\maketitle

Facultad de Matem\'aticas, Pontificia Universidad Cat\'olica de Chile, 

Vicu\~na Mackenna 4860, Santiago de Chile 

\bigskip

E-mail: {\it disambou@mat.uc.cl}

\begin{abstract} 
We consider Dirac, Pauli and Schr\"odinger quantum magnetic Hamiltonians of full rank in ${\rm L}^2 \big( {\br^{2d}} \big)$, $d \ge 1$, perturbed 
by non-self-adjoint (matrix-valued) potentials. 
On the one hand, we show the existence of non-self-adjoint perturbations, generating near each point of the essential spectrum of the 
operators, infinitely many (complex) eigenvalues. In particular, we establish point spectrum analogous of B\"{o}gli results \cite{bog} 
obtained for non-magnetic Laplacians, and hence showing that classical Lieb-Thirring inequalities cannot hold for our magnetic models. 
On the other hand, we give asymptotic behaviours of the number of the (complex) eigenvalues. In particular, for compactly supported potentials, 
our results establish non-self-adjoint extensions of Raikov-Warzel \cite{raik} and Melgaard-Rozenblum \cite{mel} results. So, we show 
how the (complex) eigenvalues converge to the points of the essential spectrum asymptotically, i.e., up to a multiplicative explicit constant, as
$$
\frac{1}{d!} \Bigg( \frac{\vert \ln r \vert}{\ln \vert \ln r \vert} \Bigg)^d, \quad r \searrow 0,
$$
in small annulus of radius $r > 0$ around the points of the essential spectrum.
\end{abstract}

\bigskip


\bigskip



\section{Introduction}\label{s0}

\subsection{Models}
In $\br^2$, consider Dirac, Pauli and Schr\"{o}dinger quantum Hamiltonians, described below, see Subsections \ref{ss1,1} and \ref{ss1,2}, with constant 
magnetic field of strength $b > 0$. To simplify the presentation, we shall not include any physical parameters. Namely, the particle mass, the particle charge, 
the speed of light, or the Planck constant are chosen equal to one. We denote $x = (x_1,x_2)$ the variables in $\br^2$, and the magnetic field $b$ is generated 
by the magnetic vector potential
\begin{equation}
{\bf A} = {\bf A} (x) = \frac{b}{2}(-x_2,x_1), \quad {\rm i.e.,} \quad b = {\rm curl} \, {\bf A}. 
\end{equation}
Let us recall and fix some useful definitions and notations. Let $M$ be a closed operator acting on a separable Hilbert space $\mathscr{H}$. An isolated point 
$\lambda$ in $\sigma(M)$, the spectrum of $M$, lies in $\sigma_{\text{\textup{disc}}}(M)$ the discrete spectrum of $M$ if it's algebraic multiplicity
$\textup{mult}(\lambda) := \textup{rank} \left( \frac{1}{2i\pi} \int_{\mathcal{C}} (M - z)^{-1}dz \right)$ is finite, $\mathcal{C}$ being a small positively oriented circle 
centred at $\lambda$ and containing $\lambda$ as the only point of $\sigma(M)$. We define the essential spectrum $\sigma_{\text{\textup{ess}}}(M)$ of $M$ as 
the set of $\lambda \in \bc$ such that $M - \lambda$ is not a Fredholm operator. When no confusion can arise in what follows below, we use the notation 
${\rm L}^2 \big( \br^2 \big) := {\rm L}^2 \big( \br^2,\bc^n \big)$ for $n = 1$, $2$, and similarly $C_0^\infty \big( \br^2 \big) := C_0^\infty \big( \br^2,\bc^n \big)$ for 
$n = 1$, $2$.

\subsubsection{Magnetic Schr\"{o}dinger operators}\label{ss1,1}

The unperturbed Schr\"{o}dinger operator $\cho(b)$ acting in ${\rm L}^2 \big( \br^2 \big)$, describes a quantum non-relativistic particle of zero spin confined to the 
$x$-plane, and subject to the magnetic field of strength $b > 0$. It is essentially self-adjoint on $C_0^\infty \big( \br^2 \big)$ and is defined by
\begin{equation}\label{Sch:ope}
\cho(b) := (-i\nabla - {\bf A})^2 - b = \Bigg( -i\frac{\partial}{\partial x_1} + \frac{bx_2}{2} \Bigg)^2 + \Bigg( -i\frac{\partial}{\partial x_2} - \frac{bx_1}{2} \Bigg)^2 - b.
\end{equation}
In the literature, the operator $\cho(b)$ is often called the Landau Hamiltonian, and it is well known that its spectrum is given by the set of the Landau levels 
({\bf LLs}) $2bq$, $q \in \mathbb{N}$, and each {\bf LL}  is an eigenvalue of infinite multiplicity. In other words, we have
\begin{equation}\label{eq:spsch}
\sigma \big( \cho(b) \big) = \sigma_{\rm ess} \big( \cho(b) \big) = \bigcup_{q=0}^{\infty} \lbrace 2bq \rbrace. 
\end{equation}
In the sequel, we set $\Lambda_q := 2bq$, $q \in \mathbb{N}$, and ${\bf P}_q$ will denote the orthogonal projection onto the eigenspace 
$\text{Ker} \, \big( \cho(b) - \Lambda_{q} \big)$. On the domain of $\cho(b)$, we define the perturbed operator 
\begin{equation}\label{Sch:opep}
\ch_V(b) := \cho(b) + V,
\end{equation}
where $V$ is the multiplication operator by the function (also) noted $V$, assumed to be complex-valued. For further use, we formulate the following 
different hypotheses on the potential $V$.

\begin{ass}\label{as1_sch}
\begin{itemize}
\item[(i)] $V$ does not vanish identically.
\item[(ii)] There exists a function $G \in {\rm L}^\infty \big( \br^{2},\br_{+}^{\ast} \big) \cap {\rm L}^{p/2} \big( \br^{2},\br_{+}^{\ast} \big)$ for some $2 \leq p < \infty$
such that $\vert V(x) \vert \leq G(x)$, $x \in  \br^2$.
\item[(iii)] $V$ is continuous on $\br^2$.
\item[(iv)] $0 \le \vert V \vert \in {\rm L}^\infty \big( \br^{2} \big)$ is mesurable, compactly supported, and $\vert V \vert > 0$ holds on an open non 
empty set of $\br^2$.
\end{itemize}
\end{ass}


\subsubsection{Magnetic Pauli and Dirac operators}\label{ss1,2}

In order to define the Pauli and Dirac operators, let us introduce the standard Pauli matrices
\begin{equation}\label{matdp}
\hat \sigma_{1} := \begin{pmatrix}
   0 & 1 \\
   1 & 0
\end{pmatrix}, \hspace{0.5cm} \hat \sigma_{2} := \begin{pmatrix}
   0 & -i \\
   i & 0
\end{pmatrix}, \hspace{0.5cm} \hat \sigma_{3} := \begin{pmatrix}
   1 & 0 \\
   0 & -1
\end{pmatrix}.
\end{equation}
The choice of the matrices $\hat \sigma_1$, $\hat \sigma_2$ and $\hat \sigma_3$ is not unique and is governed by the anti-commutation relations
\begin{equation}\label{matdpr}
 \hat \sigma_{j} \hat \sigma_{k} + \hat \sigma_{k} \hat \sigma_{j} = 2\delta_{jk} {\bf I}_2, 
\quad {\bf I}_2 := \begin{pmatrix}
   1 & 0 \\
   0 & 1
\end{pmatrix},
\end{equation}
where $\delta_{jk}$ is the classical Kronecker symbol defined by $\delta_{jk} = 1$ if $j = k$, and $\delta_{jk} = 0$ for $j \neq k$. 

The unperturbed Pauli operator $P_0(b)$ acting in ${\rm L}^2 \big( \br^2 \big)$, describes a quantum non-relativistic particle of $\frac{1}{2}$-spin confined to 
the $x$-plane, and subject to the magnetic field of strength $b > 0$. It is essentially self-adjoint on $C_0^\infty \big( \br^2 \big)$ and is defined by
\begin{equation}\label{Pau:ope}
P_0(b) := \big( \hat \sigma \cdot (-i\nabla - {\bf A}) \big)^2 = (-i\nabla - {\bf A})^2 {\bf I}_2 - b \hat \sigma_3, \quad \hat \sigma := \big( \hat \sigma_1,\hat \sigma_2 \big).
\end{equation}
More explicitly, we have
\begin{equation}\label{emp}
P_0(b) = \begin{pmatrix}
   (-i\nabla - {\bf A})^2 - b & 0 \\
   0 & (-i\nabla - {\bf A})^2 + b
\end{pmatrix} = \begin{pmatrix}
   \cho(b) & 0 \\
   0 & \cho(b) + 2b
\end{pmatrix},
\end{equation}
showing, thanks to \eqref{eq:spsch}, that the spectrum of the operator $P_0(b)$ is given by the set of the Landau-Pauli levels ({\bf LPLs}) $\Lambda_q = 2bq$, 
$q \in \mathbb{N}$, with
\begin{equation}\label{eq:sppau}
\sigma \big( P_0(b) \big) = \sigma_{\rm ess} \big( P_0(b) \big) = \bigcup_{q=0}^{\infty} 
\lbrace 2bq \rbrace.
\end{equation}
In the sequel, we denote $\widetilde {\bf P}_q$ the orthogonal projection onto the eigenspace $\text{Ker} \, \big( P_0(b) - \Lambda_{q} \big)$.

The unperturbed Dirac operator $D_0(b)$ acting in ${\rm L}^2 \big( \br^2 \big)$, describes a quantum relativistic particle of $\frac{1}{2}$-spin confined to the $x$-plane, 
and subject to the magnetic field of strength $b > 0$. It is essentially self-adjoint on $C_0^\infty \big( \br^2 \big)$ and is defined by
\begin{equation}\label{Dir:ope}
D_0(b) := \hat \sigma \cdot (-i\nabla - {\bf A}) + \hat \sigma_3.
\end{equation}
Furthermore, we have the identity 
\begin{equation}\label{Dir:opecar}
D_0(b)^2 = P_0(b) + {\bf I}_2 = \begin{pmatrix}
   \cho(b) + 1 & 0 \\
   0 & \cho(b) + 2b + 1
\end{pmatrix}.
\end{equation}
It is well know that the spectrum of the operator $D_0(b)$ is given by the set of the Dirac-Landau levels ({\bf DLLs})
\begin{equation}
\Lambda_q^- := -\sqrt{2bq + 1}, \: q \in \bn^\ast, \quad {\rm and} \quad \Lambda_q^+ := \sqrt{2bq + 1}, \: q \in \bn,
\end{equation}
and each {\bf DLL} $\Lambda_q^\pm$ is an eigenvalue of infinite multiplicity. In other words, we have
\begin{equation}\label{eq:spdi}
\sigma \big( D_0(b) \big) = \sigma_{\rm ess} \big( D_0(b) \big) = 
\big\lbrace \cup_{q=1}^{\infty} \big\lbrace \Lambda_q^- \big\rbrace \big\rbrace 
\bigcup \big\lbrace \cup_{q=0}^{\infty} \big\lbrace \Lambda_q^+ \big\rbrace \big\rbrace.
\end{equation}
In the sequel, we denote ${\bf P}_q^\pm$ the orthogonal projection onto the eigenspace $\text{Ker} \, \big( D_0(b) - \Lambda_{q}^\pm \big)$. 

On the domain of the operators $P_0(b)$ and $D_0(b)$, we define the perturbed operators
\begin{equation}\label{PaDi:opep}
P_V(b) := P_0(b) + V \quad {\rm and} \quad D_V(b) := D_0(b) + V,
\end{equation}
where $V$ is the multiplication operator by the non-hermitian matrix-valued function (also) noted 
\begin{equation}\label{eqpot} 
V = \big\lbrace V_{jk}(x) \big\rbrace_{j,k=1}^2 =
 \begin{pmatrix}
   V_{11}(x) & V_{12}(x) \\
   V_{21}(x) & V_{22}(x)
\end{pmatrix} \not \equiv 0, \quad x \in  \br^2.
\end{equation}
For further use, we introduce the following different conditions on $V$ and the coefficients $V_{jk}$.

\begin{ass}\label{as1_paudi} 
\begin{itemize}
\item[(i)] $V$ does not vanish identically.
\item[(ii)] There exists a function $G \in {\rm L}^\infty \big( \br^{2},\br_{+}^{\ast} \big) \cap {\rm L}^{p/2} \big( \br^{2},\br_{+}^{\ast} \big)$ for some $2 \leq p < \infty$, 
such that $\vert V_{jk}(x) \vert \leq G(x)$, $1 \le j,k \le 2$, $x \in  \br^2$.
\item[(iii)] $V_{jk}$ is continuous on $\br^2$, $1 \le j,k \le 2$.
\item[(iv)] All the coefficients $V_{jk}$, except finitely many that vanish identically, satisfy: $0 \le \vert V_{jk} \vert \in {\rm L}^\infty \big( \br^{2} \big)$ is mesurable, 
compactly supported, and $\vert V_{jk} \vert > 0$ holds on an open non empty set of $\br^2$.
\end{itemize}
\end{ass}


\subsection{Description of our results}

Let $\mathcal{H}_V(b)$ denotes either $\ch_V(b)$, either $P_V(b)$, or $D_V(b)$. Under Assumptions \ref{as1_sch} (ii) or (iv), and Assumptions 
\ref{as1_paudi} (ii) or (iv), we establish Schatten-von Neumann bounds implying in particular that $V$ is a relatively compact perturbation w.r.t. the 
operator $\mathcal{H}_{0}(b)$, see Propositions \ref{l_estsch}, \ref{l_estpau} and \ref{l_estdi} respectively. Thus, the Weyl criterion on the invariance of 
the essential spectrum implies that $\sigma_{\textup{ess}} \big( \mathcal{H}_V(b) \big) = \sigma_{\textup{ess}} \big( \mathcal{H}_0(b) \big)$. However, 
\cite[Theorem 2.1, p. 373]{goh} implies that the operator $\mathcal{H}_V(b)$ can have a discrete spectrum 
$\sigma_{\textup{disc}} \big( \mathcal{H}_V(b) \big)$ that can only accumulate at $\sigma_{\textup{ess}} \big( \mathcal{H}_0(b)  \big)$ given by the set 
of the Dirac-Landau-Pauli levels ({\bf DLPLs}). Presently, the spectral analysis of non-self-adjoint quantum Hamiltonians is widely addressed, and, recently, 
accumulation problems on complex eigenvalues are investigated by several authors in various (non-self-adjoint) situations, see for instance the articles \cite{alm,bog,cue,eng,pav,dio1-17,dio2-17,wan} and the references cited there. It is well known, see for instance \cite{ra,mel} (see also the references therein), 
that when the operators $\mathcal{H}_0(b)$ are perturbed by self-adjoint electric potentials, then, accumulation of (real) discrete eigenvalues can happen 
near each point of their essential spectrum. 
However, as far we know, there are no such results when they are perturbed by non-self-adjoint electric potentials. The purpose of this paper is to try to 
fill this gap by announcing and giving an overview of new results in this direction. In particular, asymptotics of the counting function of the complex eigenvalues 
are obtained.
More precisely, in a small annulus $\Omega_q(a_1,a_2) := \lbrace \lambda \in \mathbb{C} : a_1 < \vert 
\Lambda_q^\# - \lambda \vert < a_2 \rbrace$ near a fixed {\bf DLPL} $\Lambda_q^\#$, $q \ge 0$, we prove, see Theorems \ref{t1,2}, \ref{t1,3}, \ref{t1,4}, the 
existence of the limit 
\begin{equation}
\lim_{r \searrow 0} \frac{\# \sigma_{\textup{disc}} \big( \mathcal{H}_{V_\omega}(b) \big) \cap \Omega_q(|\omega|r,|\omega|r_0)}
{\textup{Tr} \, \textbf{\textup{1}}_{[r,\infty)} \big( {\bf P}_{q}^\# \vert W \vert {\bf P}_{q}^\# \big)},
\end{equation}
for some oriented potentials $V_\omega = \omega W$, $\omega \in \bc^\ast$, with $W$ of definite sign,
and where ${\bf P}_q^\#$ denotes the orthogonal projection onto the eigenspace associated with the eigenvalue $\Lambda_q^\#$. As consequence, we derive 
from our main asymptotics results, magnetic analogous, see Theorems \ref{tc1,1}, \ref{tc1,2}, \ref{tc1,3} and their generalizations, of the following recent results 
by B\"{o}gli established for non-magnetic Laplace operators:

\begin{theo}{\cite[Theorem 1]{bog}}\label{tb1}
Let $p > d \ge 1$ and $\mathcal{E} > 0$. There exists $V \in {\rm L}^\infty \big( \br^d \big) \cap {\rm L}^p \big( \br^d \big)$ with 
$\max \big\lbrace \Vert V \Vert_\infty,\Vert V \Vert_{{\rm L}^p} \big\rbrace \le \mathcal{E}$ that decays at infinity so that the Schr\"{o}dinger operator $H := -\Delta + V$, 
$\mathcal{D}(H) := W^{2,2} \big( \br^d \big)$, has infinitely many eigenvalues in the open lower complex half-plane that accumulate at every point in $[0,\infty)$.
\end{theo}

\noindent
Set $\br_+^d := \big\lbrace (x_1,\ldots,x_d) \br^d : x_d > 0 \big\rbrace$ and impose (real) Robin boundary conditions.

\begin{theo}{\cite[Theorem 2]{bog}}\label{tb2}
Let $p > d \ge 1$ and $\mathcal{E} > 0$, and let $\phi \in [0,\pi)$. There exists $V \in {\rm L}^\infty \big( \br_+^d \big) \cap {\rm L}^p \big( \br_+^d \big)$ with 
$\max \big\lbrace \Vert V \Vert_\infty,\Vert V \Vert_{{\rm L}^p} \big\rbrace \le \mathcal{E}$ that decays at infinity so that the Schr\"{o}dinger operator $H := -\Delta + V$, 
$\mathcal{D}(H) := \big\lbrace f \in W^{2,2} \big( \br_+^d \big) : \cos(\phi) \partial_{x_d} f + \sin(\phi) f = 0 \: {\rm on} \: \partial \br_+^d \big\rbrace$, has infinitely many 
eigenvalues in the open lower complex half-plane that accumulate at every point in $[0,\infty)$.
\end{theo}

\noindent
In particular, for $V$ compactly supported, our results establish non-self-adjoint extensions of Raikov-Warzel \cite[Theorem 2.2]{raik} and Melgaard-Rozenblum 
\cite[Theorems 1.2 and 1.3]{mel}, showing how the (complex) eigenvalues converge to the {\bf DPLLs} asymptotically, see Remarks \ref{rschb} (b) and 
\ref{rdb} (b), together with their generalizations \eqref{re:asbg} and \eqref{re:asdg}.
In comparison with B\"ogli results, note that the nature of our accumulation phenomena is closely related to the degeneration of the {\bf DPLLs},
which is characterized by the preponderance role of the Toeplitz operators ${\bf P}_{q}^\# \vert W \vert {\bf P}_{q}^\#$.
A key ingredient of the proof of our results is powerful theoretical recent results established in \cite{bo}.
Otherwise, it is also interesting to mention the following fact: the classical Lieb-Thirring inequalities could be interpreted as a bridge between quantum and 
classical mechanics, having important applications in the mathematical theory of stability of matter. If we consider an appropriate decaying potential $V : \mathbb{R}^d \longrightarrow \mathbb{R}$, $d \geq 2$, with a non trivial negative part, 
and consider $\sigma_{{\rm disc}}(-\Delta + V)$ the discrete spectrum (namely the set of negative eigenvalues counted with the multiplicities) of the self-adjoint 
Schr\"odinger operator $-\Delta + V$, then, the classical Lieb-Thirring inequalities, see \cite{lt1} for the original work, read
\begin{equation}\label{eqlt}
\sum_{\lambda \, \in\, \sigma_{{\rm disc}}(-\Delta + V)} \vert \lambda \vert^\gamma \leq C(\gamma,d) \int_{\mathbb{R}^d} V(x)_-^{\gamma + d/2} dx,
\end{equation}
with appropriate $\gamma \geq 0$, and a constant $C(\gamma,d) > 0$ which depends only on $\gamma$ and $d$. Theorems \ref{tc1,1}, \ref{tc1,2},
\ref{tc1,3} and their generalizations below, point out in particular the existence of non-self-adjoint perturbations $V$ for which each element of 
$\sigma_{\textup{ess}} \big( \mathcal{H}_V(b) \big)$ is an accumulation point of a sequence of complex eigenvalues lying in $\sigma_{\textup{disc}} 
\big( \mathcal{H}_V(b) \big)$. Therefore, this implies that the Lieb-Thirring inequality \eqref{eqlt} cannot be satisfied in this case for the operators 
$\mathcal{H}_V(b)$.

Our paper is organized as follows. In Section \ref{smr}, we formulate our mains results. In Section \ref{s003}, we establish preliminary Schatten-von 
Neumann bounds we need on the free operators. In Section \ref{s03}, we reduce our problem to the analysis of zeros of holomorphic regularized
determinant functions. Section \ref{s4} is devoted to the proof our main results.

\section{Main results}\label{smr}

{\bf Notations.} We adopt mathematical physics and spectral analysis notations and terminologies from Reed-Simon \cite{ree}. Recall that a compact operator $K$, 
i.e. $K \in \sinf$, defined on a separable Hilbert space belongs to the Schatten-von Neumann class ideals $\spp, p \ge 1$, if 
\begin{equation}
\Vert K \Vert_{\spp} := \big( {\rm Tr} \, \vert K \vert^{p} \big)^{1/p} < \infty.
\end{equation}
We refer the reader to Simon \cite{sim} and Gohberg-Goldberg-Krupnik \cite{gohb} for further information on the subject. In the sequel, as usual, 
the resolvent set of an operator $M$ will be denoted $\rho(M)$.

\subsection{Results on Schr\"{o}dinger operators}

We shall consider the following class of non-self-adjoint perturbations:

\begin{ass}\label{as2_sch}
$V$ is a complex-valued potential of the form $V = V_\omega := \omega W$ with $\omega \in \bc$ and $W$ a real-valued potential such that $\pm W \ge 0$.
\end{ass}

We recall that ${\bf P}_q$, $q \ge 0$, defines the orthogonal projection onto  $\text{Ker} \, \big( H_0(b) - \Lambda_{q} \big)$ for a given {\bf LL} 
$\Lambda_q = 2bq$. Let $V$ satisfy Assumptions \ref{as1_sch} (ii)-(iii) and \ref{as2_sch}, or Assumptions \ref{as1_sch} (iv) and \ref{as2_sch}.
Firstly, this implies that $\sqrt{\vert W \vert} {\bf P}_q$ is compact for any $q \ge 0$. To see this, consider for instance the formula 
$(\cho(b) - \lambda)^{-1} = \sum_{q \ge 0} {\bf P}_q (\Lambda_q - \lambda)^{-1}$ for $\lambda \in \rho \big( \cho(b) \big)$, and observe that 
\begin{equation}\label{eq:tosch}
\sqrt{\vert W \vert} {\bf P}_q = (\Lambda_q - \lambda) \sqrt{\vert W \vert} \big( \cho(b) - \lambda \big)^{-1} {\bf P}_q \in \spp \subset \sinf,
\end{equation} 
by Proposition \ref{l_estsch} (see also \cite{dr}). Secondly, \cite[Proposition 7.1]{mel} or \cite[Lemma 3.5]{raik}  implies that
${\rm rank} \, \big( \sqrt{\vert W \vert} {\bf P}_q \sqrt{\vert W \vert} \big) = {\rm rank} \, \big( {\bf P}_q \vert W \vert {\bf P}_q \big) = \infty$. 
In the sequel, our results will be closely related to the Toeplitz operator ${\bf P}_{q} \vert W \vert {\bf P}_{q}$, $q \ge 0$.
Near a fixed {\bf LL} $\Lambda_q = 2bq$, $q \ge 0$, the eigenvalues of the operator $\ch_V(b)$ can be parametrized by 
$\lambda_q = \lambda_{q}(k) := \Lambda_q - k$, with $k$ small enough, see Section \ref{s03} for more details. For $s_{0}$, $\delta$ two positive constants fixed 
and $s > 0$ tending to zero, we define the sector
\begin{equation}\label{eq1,10}
{\mathcal S}(\delta,s,s_{0}) := \big\lbrace x + iy \in \mathbb{C} : s < x < s_{0}, -\delta x < y < \delta x \big\rbrace,
\end{equation}
and the counting function 
\begin{equation}
\mathcal{N}_{q,\ch_V(b)}(s,s_0) := \# \Big\lbrace \lambda_{q}(k) \in \sigma_{\textup{disc}} \big( \ch_{V}(b) \big) : s < \vert k \vert < s_{0} \Big\rbrace.
\end{equation}

\begin{theo}\label{t1,2}
Let $V = V_\omega$ satisfy Assumptions \ref{as1_sch} (i)-(ii)-(iii) and \ref{as2_sch}, or Assumptions \ref{as1_sch} (iv) and \ref{as2_sch}. 
Fix a {\bf LL}  $\Lambda_q = 2bq$. Then, there exists a discrete set $\Sigma_q \subset \bc^\ast$ such that for all 
$\omega \in \bc^\ast \setminus \Sigma_q$, the operator $\ch_{V_\omega}(b)$ satisfies the following: there exists $r_{0} > 0$ such that:
\begin{itemize}
\item[(i)] $\lambda_q = \lambda_q(k) \in \sigma_{\textup{disc}} \big( \ch_{V_\omega}(b) \big)$, $\vert \omega \vert r < \vert k \vert < \vert \omega \vert r_{0}$, 
satisfies
\begin{equation}\label{eq1,11}
\lambda_q \in \Lambda_q \, \pm \omega \overline{{\mathcal S}(\delta,r,r_{0})}, \quad
\delta > 0.
\end{equation}

\item[(ii)] The number of eigenvalues of $\ch_{V_\omega}(b)$ near $\Lambda_q$ is infinite. Moreover, there exists a sequence 
$(r_{\ell})_{\ell}$ of positive numbers tending to zero such that 
\begin{equation}\label{eq1,12}
\lim_{\ell \longrightarrow \infty} \frac{\mathcal{N}_{q,\ch_{V_\omega}(b)} \big( \vert \omega \vert r_\ell,\vert \omega \vert r_0 \big)}
{\textup{Tr} \, \textbf{\textup{1}}_{[r_{\ell},\infty)} \big( {\bf P}_{q} \vert W \vert {\bf P}_{q} \big)} = 1.
\end{equation}
\end{itemize}
\end{theo}

\begin{rem}\label{rsch}
\begin{itemize}
\item[(a)] Theorem \ref{t1,2} remains valid if the condition $\omega \in \bc^\ast \setminus \Sigma_q$ is replaced by $\omega$ small enough.

\item[(b)] When the function $\vert W \vert : \br^2 \rightarrow \br_+$ admits a power-like decay, an exponential decay, or is compactly supported, 
then, asymptotic behaviours of $\textup{Tr} \, {\bf 1}_{[r,\infty)} \big( {\bf P}_q \vert W \vert {\bf P}_q \big)$ as $r \searrow 0$ are well known from 
\cite[Theorem 2.6]{ra}, \cite[Lemma 3.4]{raik} and \cite[Lemma 3.5]{raik}, respectively. In particular, such asymptotics show that 
$\textup{Tr} \, {\bf 1}_{[r,\infty)} \big( {\bf P}_q \vert W \vert {\bf P}_q \big) \rightarrow \infty$ as $r \searrow 0$. In this case, in Theorem \ref{t1,2}, the 
eigenvalues of the operator $\ch_{V_\omega}(b)$ satisfy near the {\bf LL} $\Lambda_q = 2bq$,
\begin{equation}\label{re:as}
\lim_{r \searrow 0} \frac{\mathcal{N}_{q,\ch_{V_\omega}(b)} \big( \vert \omega \vert r,\vert \omega \vert r_0 \big)}{\textup{Tr} \, \textbf{\textup{1}}_{[r,\infty)} \big( {\bf P}_{q} 
\vert W \vert {\bf P}_{q} \big)} = 1.
\end{equation}
\end{itemize}
\end{rem}

\noindent
A consequence of Theorem \ref{t1,2} is the following result:

\begin{theo}\label{tc1,1}
Let $p \ge 2$. Then, there exists a complex-valued potential $V \in {\rm L}^\infty \big( \br^2 \big) \cap {\rm L}^{p/2} \big( \br^2 \big)$ decaying at infinity, 
generating near each {\bf LL} $\Lambda_q = 2bq$, $q \ge 0$, infinitely many eigenvalues lying in $\sigma_{\rm disc}(\ch_V(b))$ that accumulate at 
$\Lambda_q$. Furthermore, they are located near a semi-axis.
\end{theo}

\noindent
\begin{prof}
According to Theorem \ref{t1,2}, it suffices to consider any potential $V = V_\omega$ satisfying Assumptions \ref{as1_sch} (i)-(ii)-(iii) and \ref{as2_sch}, 
decaying at infinity, or Assumptions \ref{as1_sch} (iv) and \ref{as2_sch}, with $\omega \in \bc^\ast \setminus \br^\ast \cup (\cup_q^\infty \Sigma_q)$.
\end{prof}

\begin{rem}\label{rschb}
\begin{itemize}
\item[(a)] Theorem \ref{tc1,1} provides a Landau analogous of Theorems \ref{tb1} and \ref{tb2}.

\item[(b)] As shows the above proof, in Theorem \ref{tc1,1}, $V = V_\omega$ can be chosen compactly supported satisfying Assumptions \ref{as1_sch} (iv) 
and \ref{as2_sch}. In this case, according to \cite[Proposition 7.1]{mel} or \cite[Lemma 3.5]{raik} together with Remark \ref{rsch} (b), we have
\begin{equation}\label{re:asb}
\lim_{r \searrow 0} \frac{\mathcal{N}_{q,\ch_{V_\omega}(b)} \big( \vert \omega \vert r,\vert \omega \vert r_0 \big)}
{\vert \ln r \vert \big( \ln \vert \ln r \vert \big)^{-1}} = 1,
\end{equation}
showing how the (complex) eigenvalues converge to the {\bf LLs} asymptotically. So, Theorem \ref{tc1,1} can be reformulated in such a way we have a 
non-self-adjoint extension of Raikov-Warzel \cite[Theorem 2.2]{raik} and Melgaard-Rozenblum \cite[Theorem 1.2]{mel} (for $d = 2$).
\end{itemize}
\end{rem}

\medskip

\noindent
{\bf Generalization to higher dimensions:} The magnetic self-adjoint Schr\"{o}dinger operators in ${\rm L}^{2}(\br^{n})$, $n \ge 2$, 
have the form $(-i\nabla - \textbf{A})^{2}$, where $\textbf{A} := (A_{1},\ldots,A_{n})$ is a magnetic potential generating the magnetic field. By introducing the 
$1$-form $\mathcal{A} := \sum_{j=1}^{n} A_jdx_{j}$, the magnetic field $\mathcal{\textbf{B}}$ can be defined as its exterior differential. 
Namely, $\mathcal{\textbf{B}} := d \mathcal{A} = \sum_{j<\nu} B_{j\nu} dx_{j} \wedge dx_\nu$ with $B_{j\nu} := \partial_{x_j} A_\nu - \partial_{x_\nu} A_j$, $j, \nu = 1,\ldots,n$.
In the case where the $B_{j\nu}$ do not depend on $x \in \mathbb{R}^{n}$, the magnetic field can be viewed as a real antisymmetric matrix 
$\textbf{B} := \big\lbrace B_{j\nu} \big\rbrace_{j,\nu=1}^{n}$. Assume that $\textbf{B} \neq 0$, put $2d := \text{rank} \, \textbf{B}$ and $m := n - 2d = \dim \text{Ker} \, \textbf{B}$. 
Introduce $b_{1} \geq \ldots \geq b_{d} > 0$ the real such numbers that the non-vanishing eigenvalues of $\textbf{B}$ coincide with $\pm ib_j$, $j = 1,\ldots,d$. Consequently, 
in appropriate cartesian coordinates $(x_{1},y_{1},\ldots, x_{d},y_{d}) \in \mathbb{R}^{2d} = {\rm Ran} \, \textbf{B}$ and 
$z = (z_{1},\ldots,z_m) \in \mathbb{R}^m = \text{Ker} \, \textbf{B}$, $m \geq 1$, the operators $(-i\nabla - \textbf{A})^{2}$ can be written as
\begin{equation}\label{opg}
(-i\nabla - \textbf{A})^{2} = \sum_{j=1}^{d} \Bigg( \left( -i \partial_{x_j} + \frac{b_{j}y_{j}}{2} \right)^{2} + \left( -i\partial_{y_j} - \frac{b_{j}x_{j}}{2} \right)^{2} \Bigg) 
+ \sum_{\ell=1}^m \partial_{z_\ell}^2.
\end{equation}
If $m = 0$, namely when $\text{rank} \, \textbf{B} = n$, the sum with respect to $\ell$ should be omitted and we get the full rank Landau Hamiltonians
\begin{equation}\label{opg1}
H_0(b_1,\cdots,b_d) = \sum_{j=1}^{d} \Bigg( \left( -i \partial_{x_j} + \frac{b_{j}y_{j}}{2} \right)^{2} + \left( -i\partial_{y_j} - \frac{b_{j}x_{j}}{2} \right)^{2} \Bigg),
\end{equation}
defined originally on $C_{0}^{\infty} \big( \mathbb{R}^{2d} \big)$. It is well known, see for instance \cite{dr,mel}, that 
$\sigma \big( H_0(b_1,\cdots,b_d) \big) = \sigma_{\textup{ess}}\big( H_0(b_1,\cdots,b_d) \big) = \cup_{q=0}^{\infty} \big\lbrace \Lambda_q \big\rbrace$, 
where the eigenvalues
\begin{equation}\label{eqnlg}
\begin{cases}
\Lambda_0 := b_1 + \cdots + b_d = \frac{1}{2} {\rm Tr} \sqrt{ {\bf B}^\ast {\bf B}}, \\ 
\Lambda_q := \inf \Big\lbrace \varrho \in \br : \varrho > \Lambda_{q-1}, \varrho = \sum_{j=1}^d (2s_j - 1)b_j, \, (s_1,\ldots,s_d) \in \bn_\ast^d \Big\rbrace, \, q \ge 1,
\end{cases}
\end{equation}
are known as the {\bf LLs}. In the particular case $b_1 = \cdots = b_d = b$, the {\bf LLs} take the more simplest form $\Lambda_q = 2b(d + 2q)$, 
$q \ge 0$. The Schr\"{o}dinger operator $\cho(b)$ defined by \eqref{Sch:ope} we consider corresponds the the case $d = 1$ with $b_1 = b$ shifted by 
$-b$. Nevertheless, in view of \cite[Proposition 7.1]{mel}, which is an extension of \cite[Lemma 3.5]{raik} to higher dimensions $2d$, $d \ge 1$, Theorems \ref{t1,2} 
and \ref{tc1,1} remain valid for the general Schr\"{o}dinger operators of full rank in  ${\rm L}^2 \big( \br^{2d} \big)$, $d \ge 1$, defined by \eqref{opg1}. 
More precisely:
\begin{itemize}
  \item[1)] In Assumptions \ref{as1_sch} (ii)-(iii)-(iv), $\br^2$ should be replaced by $\br^{2d}$.
  \item[2)] In Theorems \ref{t1,2} and \ref{tc1,1}, $p$ should satisfy $p \ge 2$ for $d = 1$ and $p > d$ for $d > 1$.
  Actually, the condition $p \ge 2$ for $d = 1$ and $p > d$ for $d > 1$ above, is the one we need to impose to get the analogous of Proposition \ref{l_estsch} 
  in the general case.
  \item[3)] In Theorem \ref{tc1,1}, the complex-valued potential $V$ should satisfy $V \in {\rm L}^\infty \big( \br^{2d} \big) \cap {\rm L}^{p/2} \big( \br^{2d} \big)$.
  \item[4)] In \eqref{eqnlg}, the number $\varkappa$ of different sets $(s_1,\ldots,s_d) \in \bn_\ast^d$ which determine one and the same {\bf LL} $\Lambda_q$
  is called the multiplicity of $\Lambda_q$. In this case, in Remark \ref{rschb} (b), according to \cite[Proposition 7.1]{mel}, \eqref{re:asb} will take the more 
  general form
\begin{equation}\label{re:asbg}
\mathcal{N}_{q,\ch_{V_\omega}(b)} \big( \vert \omega \vert r,\vert \omega \vert r_0 \big) \sim
\varkappa \frac{1}{d!} \Bigg( \frac{\vert \ln r \vert}{\ln \vert \ln r \vert} \Bigg)^d, \quad r \searrow 0.
\end{equation}
\end{itemize} 

\subsection{Results on Pauli and Dirac operators}

We conserve the notations introduced previously. 
As above, we need to put an additional assumption on the matrix perturbation $V$ as follows:

\begin{ass}\label{as2_padi}
$V$ is a matrix-valued potential of the form $V = V_\omega := \omega W$, with $\omega \in \bc$, and
$
W = \begin{pmatrix}
   W_{11}(x) & W_{12}(x) \\
   W_{21}(x) & W_{22}(x)
\end{pmatrix}
$
is hermitian such that $\pm W \ge 0$ in the form sense.
\end{ass}

\subsubsection{The Pauli case}

Note that the matrix $\vert W \vert$ satisfies $\vert W \vert = \pm W$ for $\pm W \ge 0$. We recall that $\widetilde {\bf P}_q$, $q \ge 0$, denotes the orthogonal 
projection onto $\text{Ker} \, \big( P_0(b) - \Lambda_{q} \big)$ for a given {\bf LPL} $\Lambda_q = 2bq$. Thus, for $V$ satisfying Assumptions \ref{as1_paudi} 
(ii)-(iii) and \ref{as2_padi}, or Assumptions \ref{as1_paudi} (iv) and \ref{as2_padi}, we have
\begin{equation}\label{eq:topau}
\sqrt{\vert W \vert} \widetilde {\bf P}_q = (\Lambda_q - \lambda) \sqrt{\vert W \vert} 
\big( P_0(b) - \lambda \big)^{-1} \widetilde {\bf P}_q \in \spp \subset \sinf,
\end{equation} 
by Proposition \ref{l_estpau}, for $\lambda \in \rho \big( P_0(b) \big)$. Moreover, since 
\begin{equation}\label{eqpq}
\widetilde {\bf P}_0 = \begin{pmatrix}
   {\bf P}_0 & 0 \\
   0 & 0
\end{pmatrix} \qquad {\rm and} \qquad 
\widetilde {\bf P}_q = \begin{pmatrix}
   {\bf P}_q & 0 \\
   0 & {\bf P}_{q - 1}
\end{pmatrix}, \: q \ge 1,
\end{equation}
${\bf P}_q$, $q \ge 0$, being the orthogonal projection onto  $\text{Ker} \, \big( H_0(b) - \Lambda_{q} \big)$, then, we have
$$ \widetilde {\bf P}_0 \vert W \vert \widetilde {\bf P}_0 = 
 \begin{pmatrix}
   {\bf P}_0 & 0 \\
   0 & 0
\end{pmatrix} \vert W \vert \begin{pmatrix}
   {\bf P}_0 & 0 \\
   0 & 0
\end{pmatrix} = \begin{pmatrix}
   \pm {\bf P}_0 W_{11} {\bf P}_0 & 0 \\
   0 & 0
\end{pmatrix} = \begin{pmatrix}
   {\bf P}_0 \vert W_{11} \vert {\bf P}_0 & 0 \\
   0 & 0
\end{pmatrix},$$
so that 
$$
{\rm rank} \, \Big( \sqrt{\vert W \vert} \widetilde {\bf P}_0 \sqrt{\vert W \vert} \Big) = {\rm rank} \, \Big( \widetilde {\bf P}_0 \vert W \vert \widetilde {\bf P}_0 \Big) = 
{\rm rank} \, \big( {\bf P}_0 \vert W_{11} \vert {\bf P}_0 \big) = \infty,
$$ 
due to \cite[Proposition 7.1]{mel} or \cite[Lemma 3.5]{raik}. Our results will be closely related to the Toeplitz operator 
$\widetilde {\bf P}_q \vert W \vert \widetilde {\bf P}_q$, $q \ge 0$. 
Near a fixed {\bf LPL} $\Lambda_q = 2bq$, $q \ge 0$, the eigenvalues of the operator $P_V(b)$ can be parametrized by 
$\lambda_q = \lambda_{q}(k) := \Lambda_q - k$, with $k$ small enough, see Section \ref{s03} for more details. As above, we define the counting function 
\begin{equation}
\mathcal{N}_{q,P_V(b)}(s,s_0) := \# \Big\lbrace \lambda_{q}(k) \in 
\sigma_{\textup{disc}} \big( P_{V}(b) \big) : s < \vert k \vert < s_{0} \Big\rbrace.
\end{equation}
Under the above considerations, we establish the following theorem:

\begin{theo}\label{t1,3}
Let $V = V_\omega$ satisfy Assumptions \ref{as1_paudi} (i)-(ii)-(iii) and \ref{as2_padi}, or Assumptions \ref{as1_paudi} (iv) and \ref{as2_padi}. 
Fix a {\bf LPL} $\Lambda_q = 2bq$. Then, there exists a discrete set $\Xi_q \subset \bc^\ast$ such that for all 
$\omega \in \bc^\ast \setminus \Xi_q$, the operator $P_{V_\omega}(b)$ 
satisfies the following: there exists $r_{0} > 0$ such that:
\begin{itemize}
\item[(i)] $\lambda_q = \lambda_q(k) \in \sigma_{\textup{disc}} \big( P_{V_\omega}(b) \big)$, $\vert \omega \vert r < \vert k \vert < \vert \omega \vert r_{0}$, 
satisfies 
\begin{equation}
\lambda_q \in \Lambda_q \, \pm \omega \overline{{\mathcal S}(\delta,r,r_{0})}, \quad \delta > 0,
\end{equation}
${\mathcal S}(\delta,r,r_{0})$ being the sector defined by \eqref{eq1,10}.

\item[(ii)] If $q = 0$, the number of eigenvalues of $P_{V_\omega}(b)$ near $\Lambda_0$ is infinite. Furthermore, there exists a positive 
sequence $(\mu_{\ell})_{\ell}$ tending to zero such that 
\begin{equation}
\lim_{\ell \longrightarrow \infty} \frac{\mathcal{N}_{q,P_{V_\omega}(b)} \big( \vert \omega \vert \mu_\ell,\vert \omega \vert r_0 \big)}{\textup{Tr} \, \textbf{\textup{1}}_{[\mu_{\ell},\infty)} \big( {\bf P}_0 \vert W_{11} \vert {\bf P}_0 \big)} = 1.
\end{equation}

\item[(iii)] If $q \ge 1$, suppose moreover that ${\rm rank} \, \Big( \widetilde {\bf P}_q \vert W \vert \widetilde {\bf P}_q \Big) = \infty$. Then,
the number of eigenvalues of $P_{V_\omega}(b)$ near $\Lambda_q$ is infinite. 
Furthermore, there exists a positive sequence $(\nu_{\ell})_{\ell}$ tending to zero such that 
\begin{equation}\label{eq:asp}
\lim_{\ell \longrightarrow \infty} \frac{\mathcal{N}_{q,P_{V_\omega}(b)} \big( \vert \omega \vert \nu_\ell,\vert \omega \vert r_0 
\big)}{\textup{Tr} \,\textbf{\textup{1}}_{[\nu_{\ell},\infty)} \Big( \widetilde {\bf P}_q \vert W \vert \widetilde {\bf P}_q \Big)} 
= 1. 
\end{equation}
\end{itemize}
\end{theo}

\begin{rem}
\begin{itemize}
\item[(a)] Theorem \ref{t1,3} remains valid if the condition $\omega \in \bc^\ast \setminus \Xi_q$ is replaced by $\omega$ small enough.
\item[(b)] Remark \ref{rsch} (b) remains valid with $\vert W \vert$ replaced by $\vert W_{11} \vert$ and the projection ${\bf P}_q$ by ${\bf P}_0$.
\end{itemize}
\end{rem}

Now, let $V$ satisfy Assumptions \ref{as1_paudi} (i)-(ii)-(iii) and \ref{as2_padi}, or Assumptions \ref{as1_paudi} (iv) and \ref{as2_padi}, with 
\begin{equation}
W = {\rm Diag}(W_{11},W_{22}) := 
\begin{pmatrix}
   W_{11}(x) & 0 \\
   0 & W_{22}(x)
\end{pmatrix}.
\end{equation}
Then, \eqref{eqpq} implies for $q \ge 1$ that
$$
\widetilde {\bf P}_q \vert W \vert \widetilde {\bf P}_q = 
 \begin{pmatrix}
   {\bf P}_q & 0 \\
   0 & {\bf P}_{q-1}
\end{pmatrix} \vert W \vert \begin{pmatrix}
   {\bf P}_q & 0 \\
   0 & {\bf P}_{q-1}
\end{pmatrix} = \begin{pmatrix}
   {\bf P}_q \vert W_{11} \vert {\bf P}_q & 0 \\
   0 & {\bf P}_{q-1} \vert W_{22} \vert {\bf P}_{q-1}
\end{pmatrix}.
$$
Thus, as above, we have ${\rm rank} \, \Big( \sqrt{\vert W \vert} \widetilde {\bf P}_q \sqrt{\vert W \vert} \Big) = 
{\rm rank} \, \Big( \widetilde {\bf P}_q \vert W \vert \widetilde {\bf P}_q \Big) = \infty$, since $${\rm rank} \, \big( {\bf P}_q \vert W_{11} \vert {\bf P}_q \big) +
{\rm rank} \, \big( {\bf P}_{q-1} \vert W_{22} \vert {\bf P}_{q-1} \big) = \infty.$$ 
Therefore, this together with Theorem \ref{t1,3} (iii) give the following corollary:

\begin{cor}\label{cp}
Under the assumptions and the notations of Theorem \ref{t1,3}, assume moreover that $W = {\rm Diag}(W_{11},W_{22})$. Then, for 
$\omega \in \bc^\ast \setminus \Xi_q$ ($q \ge 1$), the number of eigenvalues of $P_{V_\omega}(b)$ near the fixed {\bf LPL} $\Lambda_q$ is infinite, and, 
there exists a positive sequence $(\nu_{\ell})_{\ell}$ tending to zero such that $\mathcal{N}_{q,P_{V_\omega}(b)} \big( \vert \omega \vert \nu_\ell,\vert 
\omega \vert r_0 \big)$ satisfies \eqref{eq:asp}.
\end{cor}

\noindent
A consequence of Theorem \ref{t1,3} (i)-(ii) and Corollary \ref{cp} is the following result:

\begin{theo}\label{tc1,2}
Let $p \ge 2$. Then, there exists a non-hermitian matrix-valued potential $V = \big\lbrace V_{jk}(x) \big\rbrace_{j,k=1}^2$, with $V_{jk} \in {\rm L}^\infty 
\big( \br^2 \big) \cap {\rm L}^{p/2} \big( \br^2 \big)$ decaying at infinity, generating near each {\bf LPL} $\Lambda_q = 2bq$, $q \ge 0$, 
infinitely many eigenvalues lying in $\sigma_{\rm disc}(P_V(b))$ that accumulate at $\Lambda_q$. Furthermore, they 
are located near a semi-axis.
\end{theo}

\noindent
\begin{prof}
Thanks to Theorem \ref{t1,3} (i)-(ii) and Corollary \ref{cp}, it suffices to consider any matrix-valued potential 
$V = V_\omega = {\rm Diag}(\omega W_{11},\omega W_{22})$, $\omega \in \bc^\ast \setminus \br^\ast \cup (\cup_q^\infty \Xi_q)$, satisfying Assumptions 
\ref{as1_paudi} (i)-(ii)-(iii) and \ref{as2_padi}, with $W_{jj}$, $j = 1$, $2$ decaying at infinity, or Assumptions \ref{as1_paudi} (iv) and \ref{as2_padi}.
\end{prof}

\begin{rem}\label{rp}
\begin{itemize}
\item[(a)] Notice that Theorem \ref{tc1,2} provides a magnetic Pauli analogous of Theorems \ref{tb1} and \ref{tb2}.

\item[(b)] The above proof shows that in Theorem \ref{tc1,2}, $V = V_\omega = {\rm Diag}(\omega W_{11},\omega W_{22})$ can be chosen 
such that $W_{jj}$, $j = 1$, $2$, satisfy Assumptions \ref{as1_paudi} (iv) and \ref{as2_padi}. In this case, if $W_{22}$ vanishes identically,
then, \eqref{re:asb} holds with $\ch_{V_\omega}(b)$ replaced by $P_{V_\omega}(b)$.
\end{itemize}
\end{rem}

\medskip

\noindent
{\bf Generalization to higher dimensions:} Let $H_0(b_1,\ldots,b_d)$, $d \ge 1$, be the Schr\"{o}dinger operators defined by \eqref{opg1}, and 
${\bf I}_{2^{d}}$ denotes the $2^d \times 2^d$ identity matrix. Then, see \cite{shi} and \cite[Identity (4.12)]{mel}, the Pauli operators of full 
rank essentially self-adjoint in  ${\rm L}^2 \big( \mathbb{R}^{2d},\mathbb{C}^{2^d} \big)$, $d \ge 1$, are originally defined on 
$C_{0}^{\infty} \big( \mathbb{R}^{2d},\mathbb{C}^{2^d} \big)$ by
\begin{equation}\label{opgp}
P_0(b_1,\ldots,b_d) = H_0(b_1,\ldots,b_d) {\bf I}_{2^{d}} - \Delta(b_1,\ldots,b_d),
\end{equation}
$\Delta(b_1,\ldots,b_d)$ being the diagonal $2^d \times 2^d$ matrix having on the diagonal the sums $\sum_{j=1}^d \varepsilon_j b_j$, where
$\varepsilon = (\varepsilon_1,\ldots,\varepsilon_d)$ belongs to the set 
$\{ (\varepsilon_1,\ldots,\varepsilon_d) : \: \textup{all possible combinations of} \: \varepsilon_j = \pm 1 \}$.
It is well-known, see \cite[Proposition 4.2]{mel}, that the spectrum of the operator $P_0(b_1,\cdots,b_d)$ is given by the eigenvalues set of the 
{\bf PLLs} with
\begin{equation}\label{eqs}
\sigma \big( P_0(b_1,\cdots,b_d) \big) = \sigma_{\textup{ess}} \big( P_0(b_1,\cdots,b_d) \big) = 
\bigg\lbrace 2 \sum_{j = 1}^d b_j  (q_j - 1) : (q_1,\ldots,q_d) \in \mathbb{N}^d \bigg\rbrace.
\end{equation}
The Pauli operator $P_0(b)$ defined by \eqref{Pau:ope} we consider corresponds the the case $d = 1$ and $b_1 = b$. 
However, in view of \cite[Proposition 7.1]{mel}, Theorems \ref{t1,3}, \ref{tc1,2} and Corollary \ref{cp} remain valid for  to the general Pauli 
operators of full rank in  ${\rm L}^2 \big( \mathbb{R}^{2d},\mathbb{C}^{2^d} \big)$, $d \ge 1$, defined by \eqref{opgp}. More precisely:
\begin{itemize}
  \item[1)] In \eqref{eqpot}, the matrix $V = \big\lbrace V_{jk}({\bf x}) \big\rbrace_{j,k=1}^{2^{d}}$ should be of size $2^d$, $d \ge 1$, 
  ${\bf x} = (x_1,y_1,\ldots,x_d,y_d) \in \br^{2d}$.
  \item[2)] In Assumptions \ref{as1_paudi} (ii)-(iii)-(iv), $\br^2$ should be replaced by $\br^{2d}$.
  \item[3)] In Theorems \ref{t1,3}, \ref{tc1,2} and Corollary \ref{cp}, $p$ should satisfy $p \ge 2$ for $d = 1$ and $p > d$ for $d > 1$.
  This condition is the one we need to impose to get the analogous of Proposition \ref{l_estpau} in the general case.
  \item[4)] In Theorem \ref{tc1,2}, the coefficients of the non-hermitian matrix-valued potential  $V$ should satisfy 
  $V_{jk} \in {\rm L}^\infty \big( \br^{2d} \big) \cap {\rm L}^{p/2} \big( \br^{2d} \big)$, $1 \le j,k \le 2^d$.
\end{itemize} 

\subsubsection{The Dirac case}

We recall that ${\bf P}_q^\pm$ denotes the orthogonal projection onto $\text{Ker} \, \big( D_0(b) - \Lambda_{q}^\pm \big)$, where $\Lambda_q^- = -\sqrt{2bq + 1}$,
$q \in \bn^\ast$, and $\Lambda_q^+ = \sqrt{2bq + 1}$, $q \in \bn$, are the {\bf DLLs}. Let $V$ satisfy Assumptions \ref{as1_paudi} (ii)-(iii) and \ref{as2_padi}, or 
Assumptions \ref{as1_paudi} (iv) and \ref{as2_padi}. Then, we have
\begin{equation}\label{eq:todi}
\sqrt{\vert W \vert} {\bf P}_q^\pm = \big( \Lambda_q^\pm - \lambda \big) \sqrt{\vert W \vert} 
\big( D_0(b) - \lambda \big)^{-1} {\bf P}_q^\pm \in \spp \subset \sinf, \quad p > 2,
\end{equation} 
by Proposition \ref{l_estdi}, for $\lambda \in \rho \big( D_0(b) \big)$. Near a fixed {\bf DLL} $\Lambda_q^\pm$, $q \ge 0$, the eigenvalues of the 
operator $D_V(b)$ can be parametrized by $\lambda_q^\pm = \lambda_{q}^\pm(k) := \Lambda_q^\pm - k$, with $k$ small enough, see Section \ref{s03} for more 
details. As above, we define the counting function 
\begin{equation}
\mathcal{N}_{q,D_V(b)}^\pm(s,s_0) := \# \Big\lbrace \lambda_{q}^\pm(k) \in \sigma_{\textup{disc}} \big( D_{V}(b) \big) : s < \vert k \vert < s_{0} \Big\rbrace,
\end{equation}
for a fixed {\bf DLL}. Under the above considerations, we establish the following theorem:

\begin{theo}\label{t1,4}
Let $V = V_\omega$ satisfy Assumptions \ref{as1_paudi} (i)-(ii)-(iii) and \ref{as2_padi}, with $p > 2$, or Assumptions \ref{as1_paudi} (iv) and \ref{as2_padi}. 
Fix a {\bf DLL} $\Lambda_q^\pm$. Then, there exists a discrete set $\Sigma_q^\pm \subset \bc^\ast$ such that for all 
$\omega \in \bc^\ast \setminus \Sigma_q^\pm$, the operator $D_{V_\omega}(b)$ satisfies the following: there exists $r_{0} > 0$ such that:

\begin{itemize}
\item[(i)] $\lambda_q^\pm = \lambda_q^\pm(k) \in \sigma_{\textup{disc}} \big( D_{V_\omega}(b) \big)$, $\vert \omega \vert r < \vert k \vert < \vert \omega \vert r_{0}$, 
satisfies 
\begin{equation}
\lambda_q^\pm \in \Lambda_q^\pm + \widetilde \omega \overline{{\mathcal S}(\delta,r,r_{0})}, 
\quad \delta > 0,
\end{equation}
where ${\mathcal S}(\delta,r,r_{0})$ is the sector defined by \eqref{eq1,10},
and $\widetilde \omega := \pm \omega$ w.r.t. $\pm W \ge 0$.

\item[(ii)] Suppose moreover that ${\rm rank} \, \big( {\bf P}_q^\pm \vert W \vert {\bf P}_q^\pm \big) = \infty$. Then, the number of eigenvalues of 
$D_{V_\omega}(b)$ near $\Lambda_q^\pm$ is infinite. Furthermore, there exists a positive sequence $(\gamma_{\ell})_{\ell}$ tending to zero such that 
\begin{equation}\label{eq:asd}
\lim_{\ell \longrightarrow \infty} \frac{\mathcal{N}_{q,D_{V_\omega}(b)}^\pm \big( \vert \omega \vert \gamma_\ell,\vert \omega \vert r_0 \big)}{\textup{Tr} \,\textbf{\textup{1}}_{[\gamma_{\ell},\infty)} \big( {\bf P}_q^\pm \vert W \vert {\bf P}_q^\pm \big)} = 1. 
\end{equation}
\end{itemize}
\end{theo}

Now, let $V$ satisfy Assumptions \ref{as1_paudi} (iv) and \ref{as2_padi} with $W = {\rm Diag}(U,U) = U {\bf I}_2$.
Then, by \cite[Proposition 8.1]{mel}, the Toeplitz operator ${\bf P}_q^\pm \vert W \vert {\bf P}_q^\pm$, $q \ge 0$, 
obeys up to a multiplicative explicit constant, the asymptotic
\begin{equation}\label{aslog}
\textup{Tr} \, \textbf{\textup{1}}_{[r,\infty)} \big( {\bf P}_q^\pm \vert W \vert 
{\bf P}_q^\pm \big) \sim \frac{\vert \ln r \vert}{\ln \vert \ln r \vert} \quad {\rm  as} 
\quad r \searrow 0.
\end{equation}
Therefore, this together with Theorem \ref{t1,4} give the following corollary:

\begin{cor}\label{cd}
Let $V = V_\omega$ satisfy Assumptions \ref{as1_paudi} (iv) and \ref{as2_padi}. Assume moreover that $W = {\rm Diag}(U,U)$. 
Then, in Theorem \ref{t1,4}, for $\omega \in \bc^\ast \setminus  \Sigma_q^\pm$, the number of eigenvalues of $D_{V_\omega}(b)$ 
near $\Lambda_q^\pm$ is infinite, and there exists a positive sequence $(\gamma_{\ell})_{\ell}$ tending to zero such that 
$\mathcal{N}_{q,D_{V_\omega}(b)}^\pm \big( \vert \omega \vert \gamma_\ell,\vert \omega \vert r_0 \big)$ satisfies \eqref{eq:asd}.
\end{cor}

\begin{rem}\label{rd}
\begin{itemize}
\item[(a)] Theorem \ref{t1,4} remains valid if the condition $\omega \in \bc^\ast \setminus \Sigma_q^\pm$ is replaced by $\omega$ small enough.
\item[(b)] In Corollary \ref{cd}, since $\vert W \vert$ is compactly supported, then, the eigenvalues of the operator 
$D_{V_\omega}(b)$ satisfy near the {\bf DLL} $\Lambda_q^\#$
\begin{equation}\label{re:asd}
\lim_{r \searrow 0} \frac{\mathcal{N}_{q,D_{V_\omega}(b)}^\pm \big( \vert \omega \vert r,\vert \omega \vert r_0 \big)}{\textup{Tr} \, \textbf{\textup{1}}_{[r,\infty)} 
\big( {\bf P}_{q}^\pm \vert W \vert {\bf P}_{q}^\pm \big)} = 1.
\end{equation}
\end{itemize}
\end{rem}

\noindent
A consequence of Theorem \ref{t1,4} and Corollary \ref{cd} is the following result:

\begin{theo}\label{tc1,3}
Let $p > 2$. Then, there exists a non-hermitian matrix-valued potential 
$V = \big\lbrace V_{jk}(x) \big\rbrace_{j,k=1}^2$, with $V_{jk} \in {\rm L}^\infty 
\big( \br^2 \big) \cap {\rm L}^{p/2} \big( \br^2 \big)$ decaying at infinity, generating near 
each {\bf DLL} $\Lambda_q^\pm$, $q \ge 0$, 
infinitely many eigenvalues lying in $\sigma_{\rm disc}(D_V(b))$ that accumulate at $\Lambda_q^\pm$. 
Furthermore, they are located near a semi-axis.
\end{theo}

\noindent
\begin{prof}
According to Theorem \ref{t1,4} and Corollary \ref{cd}, it suffices to consider any matrix-valued potential $V = V_\omega = {\rm Diag}(\omega U,\omega U)$ 
satisfying Assumptions \ref{as1_paudi} (iv) and \ref{as2_padi}, with $\omega \in \bc^\ast \setminus \br^\ast \cup (\cup_q^\infty \Sigma_q^\pm)$.
\end{prof}

\begin{rem}\label{rdb}
\begin{itemize}
\item[(a)] Theorem \ref{tc1,3} provides a magnetic Dirac analogous of Theorems \ref{tb1} and \ref{tb2}.

\item[(b)] As shows the above proof, in Theorem \ref{tc1,3}, $V$ can be chosen of the  form $V = V_\omega = {\rm Diag}(\omega U,\omega U)$, 
compactly supported satisfying Assumptions \ref{as1_paudi} (iv) and \ref{as2_padi}. In this case, according to \cite[Proposition 8.1]{mel} together with 
Remark \ref{rd} (b), we have up to a multiplicative explicit constant,
\begin{equation}\label{re:asdb}
\mathcal{N}_{q,D_{V_\omega}(b)} \big( \vert \omega \vert r,\vert \omega \vert r_0 \big)
\sim \frac{\vert \ln r \vert}{\ln \vert \ln r \vert}, \quad r \searrow 0.
\end{equation}
showing how the (complex) eigenvalues converge to the {\bf DLLs} asymptotically. Hence, Theorem \ref{tc1,3} can be reformulated in such a way we 
have a non-self-adjoint extension of Melgaard-Rozenblum \cite[Theorem 1.3]{mel} (for $d = 2$).
\end{itemize}
\end{rem}

\medskip

\noindent
{\bf Generalization to higher dimensions:} To define the Dirac operators of full rank in higher dimensions $2d$, $d \ge 1$, we refer for instance to the 
description given in \cite[Section 4]{mel} and \cite{shi} for more details. 
For a given $d \ge 1$, let $\sigma_1^{(d)},\cdots,\sigma_{2d}^{(d)},\sigma_0^{(d)}$ be the $d + 1$ Dirac matrices of size $2^{d}$, governed, as in \eqref{matdpr}, 
by the relations
\begin{equation}
\big( \sigma_j^{(d)} \big)^{\ast} = \sigma_j^{(d)} \quad {\rm and} \quad \sigma_j^{(d)} \sigma_k^{(d)} + \sigma_k^{(d)} \sigma_j^{(d)} = 2\delta_{jk} {\bf I}_{2^{d}}, 
\quad 0 \le j,k \le 2^d,
\end{equation}
where ${\bf I}_{2^{d}}$ denotes the $2^d \times 2^d$ identity matrix.. For $b_j \in \br$, $1 \le  j \le d$, $(x_1,y_1,\ldots,x_d,y_d) \in \br^{2d}$, introduce 
the operators
$P_{2j-1} = \big( -i \partial_{x_j} + \frac{b_{j}y_{j}}{2} \big)$ and $P_{2j}  = \big( -i\partial_{y_j} - \frac{b_{j}x_{j}}{2} \big)$.
Then, the Dirac operators of full rank essentially self-adjoint in  ${\rm L}^2 \big( \mathbb{R}^{2d},\mathbb{C}^{2^d} \big)$, $d \ge 1$, are originally defined on 
$C_{0}^{\infty} \big( \mathbb{R}^{2d},\mathbb{C}^{2^d} \big)$ by
\begin{equation}\label{opgd}
D_0(b_1,\ldots,b_d) = \sum_{j=1}^{2d} \sigma_j^{(d)} P_j + \sigma_0^{(d)}.
\end{equation}
It is well-known, see for instance \cite{mel}, that the spectrum of the operator $D_0(b_1,\cdots,b_d)$ is given by the eigenvalues set of the {\bf DLLs} with
\begin{equation}\label{eqs}
\sigma \big( D_0(b_1,\cdots,b_d) \big) = \sigma_{\textup{ess}} \big( D_0(b_1,\cdots,b_d) \big) = 
\big\lbrace \pm \sqrt{{\rm I}_{\textup{{\bf q}}} + 1} : {\bf q} = (q_1,\ldots,q_d) \in \mathbb{N}^d \big\rbrace, 
\end{equation}
where ${\rm I}_{\textup{{\bf q}}}$ can be expressed as ${\rm I}_{\textup{{\bf q}}}  = 2 \sum_{j = 1}^d \vert b_j \vert  (q_j - 1)$. Note that in \eqref{eqs}, the 
symmetry of $\pm \sqrt{{\rm I}_{\textup{{\bf q}}} + 1}$ breaks down for the "lowest" {\bf DLL} $\pm \sqrt{{\rm I}_0 + 1} = \pm 1$ corresponding to 
${\bf q} = (1,\ldots,1)$. It is either $1$ or $-1$. The Dirac operator $D_0(b)$ defined by \eqref{Dir:ope} we consider corresponds the the case $d = 1$ and 
$b_1 = b$. However, Theorems \ref{t1,4} remains valid for the general Dirac operators of full rank in  ${\rm L}^2 \big( \mathbb{R}^{2d},\mathbb{C}^{2^d} \big)$, 
$d \ge 1$, defined by \eqref{opgd}. Furthermore, in view of \cite[Proposition 8.1]{mel}, Corollary \ref{cd} and Theorem \ref{tc1,3} remain also valid for the Dirac 
operators \eqref{opgd}. More precisely:
\begin{itemize}
   \item[1)] In \eqref{eqpot}, the matrix $V = \big\lbrace V_{jk}({\bf x}) \big\rbrace_{j,k=1}^{2^{d}}$ should be of size $2^d$, $d \ge 1$, 
  ${\bf x} = (x_1,y_1,\ldots,x_d,y_d) \in \br^{2d}$.
  \item[2)] In Assumptions \ref{as1_paudi} (ii)-(iii)-(iv), $\br^2$ should be replaced by $\br^{2d}$.
  \item[3)] In Theorems \ref{t1,4}, \ref{tc1,3} and Corollary \ref{cd}, $p$ should satisfy $p > 2d$ for $d \ge 1$.
  The condition $p > 2d$,  $d \ge 1$ above, is the one we need to impose to get the analogous of Proposition \ref{l_estdi} in the general case.
  \item[4)] In Theorem \ref{tc1,3}, the coefficients of the non-hermitian matrix-valued potential  $V$ should satisfy 
  $V_{jk} \in {\rm L}^\infty \big( \br^{2d} \big) \cap {\rm L}^{p/2} \big( \br^{2d} \big)$, $1 \le j,k \le 2^d$.
  \item[5)] In Remark \ref{rdb} (b), according to \cite[Proposition 8.1]{mel}, \eqref{re:asdb} will take the more 
  general form
\begin{equation}\label{re:asdg}
\mathcal{N}_{{\bf q},D_{V_\omega}(b)} \big( \vert \omega \vert r,\vert \omega \vert r_0 \big) \sim
\frac{1}{d!} \Bigg( \frac{\vert \ln r \vert}{\ln \vert \ln r \vert} \Bigg)^d, \quad r \searrow 0,
\end{equation}
up to a multiplicative explicit constant given by (4.17) of \cite{mel}.
\end{itemize} 

\section{Schatten-von Neumann bounds}\label{s003}

In this section, we establish useful Schatten-von Neumann bounds implying in particular the relatively compactness 
of the potential perturbation w.r.t. the free operators. We conserve the notations introduced above.

\subsection{Bounds on Schr\"{o}dinger operators}

\begin{prop}\label{l_estsch}
\begin{itemize}
\item[(i)] Let $V$ be complex-valued satisfying Assumption \ref{as1_sch} (ii), and $\lambda \in \bc \setminus \cup_{q = 0}^\infty \lbrace \Lambda_q \rbrace$. 
Then, $\sqrt{\vert V \vert} \big( \cho(b) - \lambda \big)^{-1} \in \spp$ and there exists a constant $C = C(p,b)$ depending only on $p \ge 2$ and $b$, such that
\begin{equation}\label{est_sch}
\Big\Vert \sqrt{\vert V \vert} \big( \cho(b) - \lambda \big)^{-1} \Big\Vert_\spp \le C \big\Vert \sqrt{G} \big\Vert_{{\rm L}^p}
\Bigg( 1 + \frac{\vert \lambda + 1 \vert} {{\rm dist} \big( \lambda,\cup_{q = 0}^\infty \lbrace \Lambda_q \rbrace \big)} 
\Bigg).
\end{equation}
\item[(ii)] For $V \in {\rm L}^\infty(\br^2)$ compactly supported, for each $p \ge 2$, the same conclusion holds with $\sqrt{G}$ replaced by 
$\sqrt{\vert V \vert}$ in the r.h.s. of \eqref{est_sch}.
\end{itemize}
In particular, in both cases, $V$ is relatively compact w.r.t. the operator $\cho(b)$.
\end{prop}

\noindent
\begin{prof}
(i) Due to Assumption \ref{as1_sch} (ii), there exists a bounded operator $\mathcal{B}$ on ${\rm L}^2 \big( \br^2 \big)$ such that 
$\sqrt{\vert V \vert} = \mathcal{B} \sqrt{G}$. Thus,
$\big\Vert \sqrt{\vert V \vert} \big( \cho(b) - \lambda \big)^{-1} \big\Vert_\spp \le C \big\Vert \sqrt{G} \big( \cho(b) - \lambda \big)^{-1} \big\Vert_\spp$
for some constant $C > 0$. Since $\sqrt{G} \in {\rm L}^p \big( \br^2 \big)$, then, to show the claim, it suffices to prove that for any  
$U \in {\rm L}^p \big( \br^2 \big)$, we have the bound
\begin{equation}\label{est_sch1}
\Big\Vert U \big( \cho(b) - \lambda \big)^{-1} \Big\Vert_\spp \le C(p,b) \Vert U \Vert_{{\rm L}^p}
\Bigg( 1 + \frac{\vert \lambda + 1 \vert} {{\rm dist} \big( \lambda,\cup_{q = 0}^\infty \lbrace \Lambda_q \rbrace \big)} 
\Bigg).
\end{equation}

a) Firstly, we shall prove \eqref{est_sch1} for $p$ even. To prove the general case, we shall use an interpolation argument. Constants will 
change from an estimate to another. Let $p$ be even. 
We have
\begin{equation}\label{eq3.1}
\left\Vert U \big( \cho(b) - \lambda \big)^{-1} \right\Vert_\spp \leq \left\Vert U \big( \cho(b) + 1 \big)^{-1} 
\right\Vert_\spp \left\Vert \big( \cho(b) + 1 \big) \big( \cho(b) - \lambda \big)^{-1} 
\right\Vert.
\end{equation}
The spectral mapping theorem yields
\begin{equation}\label{eq4,10}
\left\Vert \big( \cho(b) + 1 \big) \big( \cho(b) - \lambda \big)^{-1} \right\Vert \leq 
\textup{sup}_{\varrho \in \sigma(\cho(b))} \left\vert \frac{\varrho + 1}{\varrho - \lambda} \right\vert \le
\Bigg( 1 + \frac{\vert \lambda + 1 \vert} {{\rm dist} \big( \lambda,\cup_{q = 0}^\infty \lbrace \Lambda_q \rbrace \big)} \Bigg).
\end{equation}
The diamagnetic inequality, see for instance \cite[Theorem 2.3]{avr} and \cite[Theorem 2.13]{sim}, implies that there 
exists a constant $C > 0$ such that
\begin{align}
\left\Vert U \big( \cho(b) + 1 \big)^{-1} \right\Vert_\spp & = \left\Vert U \big( (-i\nabla - {\bf A})^2 - b + 1 \big)^{-1} 
\right\Vert_\spp \nonumber \\
& \le \left\Vert U \big( (-i\nabla - {\bf A})^2 + 1 \big)^{-1} \right\Vert_\spp 
\left\Vert \big( (-i\nabla - {\bf A})^2 + 1 \big) \big( (-i\nabla - {\bf A})^2 - b + 1 \big)^{-1} \right\Vert \nonumber \\
& = \left\Vert U \big( (-i\nabla - {\bf A})^2 + 1 \big)^{-1} \right\Vert_\spp 
\left\Vert I + \big( \cho(b) + 1 \big)^{-1}b \right\Vert \nonumber \\
& \leq C \left\Vert U (-\Delta + 1)^{-1} \right\Vert_\spp C(b) = C(b) \left\Vert U (-\Delta + 1)^{-1} \right\Vert_\spp.
\label{eq4,11}
\end{align}
Now, since $p$ is even, then, by the standard criterion \cite[Theorem 4.1]{sim}, it follows that
\begin{equation}\label{eq4,12}
\big\Vert U (-\Delta + 1)^{-1} \big\Vert_\spp \leq C \Vert U \Vert_{{\rm L}^p} \left\Vert 
\Bigl( \vert \cdot \vert^{2} + 1 \Bigr)^{-1} \right\Vert_{{\rm L}^p}.
\end{equation}
Thus, estimate \eqref{est_sch1}, for $p$ even, follows by putting together bounds \eqref{eq3.1}, \eqref{eq4,10}, 
\eqref{eq4,11} and \eqref{eq4,12}. 

b) Let us show now that \eqref{est_sch1} is true for each $p \geq 2$. For any $p > 2$, 
there exists even integers $p_{0} < p_{1}$ such that $p \in (p_{0},p_{1})$ with $p_{0} \geq 2$. Let $\gamma \in (0,1)$ 
with $p = (1-\gamma)p_0 + \gamma p_1$, and consider the operator
$$
{\rm L}^{p_i} \big( \br^2 \big) \ni U \overset{M}{\longmapsto} U \big( \cho(b) - \lambda \big)^{-1} \in {\bf S}_{p_{i}}.
\quad i = 0, 1.
$$
For $i = 0$, $1$, let $C_{i} = C(p_{i},b)$ denote the constant appearing in \eqref{est_sch1}, and define
$$
C(\lambda,p_{i},b) := C_i \Bigg( 1 + \frac{\vert \lambda + 1 \vert} {{\rm dist} \big( \lambda,\cup_{q = 0}^\infty 
\lbrace \Lambda_q \rbrace \big)} \Bigg).
$$
Bound \eqref{est_sch1} implies that $\Vert M \Vert \leq C(\lambda,p_{i},b)$ for $i = 0$, $1$. By using the 
Riesz-Thorin Theorem, see for instance \cite[Sub. 5 of Chap. 6]{fol}, \cite{rie}, \cite{tho}, \cite[Chap. 2]{lun},
we can interpolate between $p_{0}$ and $p_{1}$ to obtain the extension $M : L^p \big( \br^2 \big) \longrightarrow \spp$, 
with
$$
\Vert M \Vert \leq C(\lambda,p_{0},b)^{1-\gamma} C(\lambda,p_{1},b)^{\gamma} \leq C(p,b) \Bigg( 1 + \frac{\vert \lambda + 1 \vert} 
{{\rm dist} \big( \lambda,\cup_{q = 0}^\infty \lbrace \Lambda_q \rbrace \big)} \Bigg).
$$
Therefore, for any $U \in L^p \big( \br^2 \big)$, we have
$$
\Vert M(U) \Vert_\spp \leq C(p,b) \Bigg( 1 + \frac{\vert \lambda + 1 \vert} {{\rm dist} \big( \lambda,\cup_{q = 0}^\infty 
\lbrace \Lambda_q \rbrace \big)} \Bigg) \Vert U \Vert_{L^p},
$$
or equivalently estimate \eqref{est_sch1}. 

(ii) For $V \in {\rm L}^\infty(\br^2)$ compactly supported, $\sqrt{\vert V \vert} \in {\rm L}^p \big( \br^2 \big)$ for each $p \ge 2$. Thus, the claim follows 
according to \eqref{est_sch1}. This concludes the proof of the proposition.
\end{prof}

\subsection{Bounds on Pauli and Dirac operators}

Concerning the Pauli operator, we have the following proposition:

\begin{prop}\label{l_estpau}
\begin{itemize}
\item[(i)] Let $V$ be non-hermitian matrix-valued satisfying Assumption \ref{as1_paudi} (ii), and $\lambda \in \bc \setminus \cup_{q = 0}^\infty 
\lbrace \Lambda_q \rbrace$. Then, $\sqrt{\vert V \vert} \big( P_0(b) - \lambda \big)^{-1} \in \spp$ 
and there exists a constant $C = C(p,b)$ depending only on $p \ge 2$ and $b$, such that
\begin{equation}\label{est_pau}
\Big\Vert \sqrt{\vert V \vert} \big( P_0(b) - \lambda \big)^{-1} \Big\Vert_\spp \le C \big\Vert \sqrt{G} \big\Vert_{{\rm L}^p}
\Bigg( 1 + \frac{\vert \lambda + 1 \vert} {{\rm dist} \big( \lambda,\cup_{q = 0}^\infty \lbrace \Lambda_q \rbrace \big)} 
\Bigg).
\end{equation}
\item[(ii)] Assume that all the $V_{jk} \in {\rm L}^\infty(\br^2)$ are compactly supported except finitely many that vanish identically. 
Then, for each $p \ge 2$, \eqref{est_pau} holds with $\sqrt{G}$ replaced by $e^{-\kappa \vert x \vert}$, $\kappa > 0$.
\end{itemize}
In particular, in both cases, $V$ is relatively compact w.r.t. the operator $P_0(b)$.
\end{prop}

\noindent
\begin{prof}
It is left to the reader since the use of the identity \eqref{emp} allows to mimic easily the proof of Proposition \ref{l_estsch}. Note that for the $V_{jk}$ as 
in (ii), Assumption \ref{as1_paudi} (ii) holds with $G = e^{-2\kappa \vert x \vert}$, $\kappa > 0$.
\end{prof}

For the Dirac operator, we have the following result:

\begin{prop}\label{l_estdi}
\begin{itemize}
\item[(i)] Let $V$ satisfy Assumption \ref{as1_paudi} (ii) and $\lambda \in \bc \setminus \big\lbrace \cup_{q=1}^{\infty} \big\lbrace \Lambda_q^- 
\big\rbrace \big\rbrace \bigcup \big\lbrace \cup_{q=0}^{\infty} \big\lbrace \Lambda_q^+ \big\rbrace \big\rbrace$. Then, 
$\sqrt{\vert V \vert} \big( D_0(b) - \lambda \big)^{-1} \in \spp$ and there exists a constant $C = C(p,b)$ depending only on $p > 2$ 
and $b$, such that
\begin{equation}\label{est_di}
\Big\Vert \sqrt{\vert V \vert} \big( D_0(b) - \lambda \big)^{-1} \Big\Vert_\spp \le C \big\Vert \sqrt{G} \big\Vert_{{\rm L}^p}
\Big( 1 + \big( \vert \lambda \vert + \vert \lambda \vert^2 \big) \big( 2 + C_1(\lambda) + C_2(\lambda) \big) \Big),
\end{equation}
where we have set
\begin{equation}\label{est_cdi}
C_1(\lambda) := \frac{\vert \lambda \vert^2} {{\rm dist} \big( \lambda^2,\cup_{q = 0}^\infty 
\lbrace \Lambda_q + 1 \rbrace \big)} \quad {\rm and} \quad
C_2(\lambda) := \frac{\vert \lambda \vert^2} {{\rm dist} \big( \lambda^2,\cup_{q = 0}^\infty 
\lbrace \Lambda_q + 2b + 1 \rbrace \big)},
\end{equation}
$\Lambda_q$, $q \ge 0$, being the {\bf LLs} of the Schr\"odinger operator $\cho(b)$.
\item[(ii)] Let all the coefficients $V_{jk} \in {\rm L}^\infty(\br^2)$ be compactly supported except finitely many that vanish identically. 
Then, for each $p \ge 2$, \eqref{est_di} holds with $\sqrt{G}$ replaced by $e^{-\kappa \vert x \vert}$, $\kappa > 0$.
\end{itemize}
In particular, in both cases, $V$ is relatively compact w.r.t. the operator $D_0(b)$.
\end{prop}

\noindent
\begin{prof}
Since  in the second point (ii) Assumption \ref{as1_paudi} (ii) holds with $G = e^{-2\kappa \vert x \vert}$, $\kappa > 0$, then, it suffices to prove only 
(i). Let $\lambda \in \bc \setminus \big\lbrace \cup_{q=1}^{\infty} \big\lbrace \Lambda_q^- \big\rbrace \big\rbrace \bigcup \big\lbrace 
\cup_{q=0}^{\infty} \big\lbrace \Lambda_q^+ \big\rbrace \big\rbrace$, the resolvent set of the operator $D_0(b)$. We have
\begin{equation}\label{eq4,23}
\big( D_0(b) - \lambda \big)^{-1} = D_0(b)^{-1} + \lambda \big( 1 + \lambda D_0(b)^{-1} \big) \big( D_0(b)^{2} - \lambda^{2} \big)^{-1}.
\end{equation} 
By setting 
\begin{equation}\label{eq4,24}
T_1(\lambda) := \lambda \big( 1 + \lambda D_0(b)^{-1} \big) \big( D_0(b)^{2} - \lambda^{2} \big)^{-1},
\end{equation} 
it follows from \eqref{eq4,23} that
\begin{equation}\label{eq4,25}
\sqrt{\vert V \vert} \big( D_0(b) - \lambda \big)^{-1} = \sqrt{\vert V \vert} D_0(b)^{-1} + \sqrt{\vert V \vert} T_1(\lambda).
\end{equation} 
Due to Assumption \ref{as1_paudi} (ii), there exists a bounded operator $\mathcal{B}$ on ${\rm L}^2 \big( \br^2 \big)$ 
such that $\sqrt{\vert V \vert} = \mathcal{B} \sqrt{G}$. Thus, it follows from \eqref{eq4,25} that there exists a constant
$C > 0$ such that
\begin{equation}\label{eq4,26}
\Big\Vert \sqrt{\vert V \vert} \big( D_0(b) - \lambda \big)^{-1} \Big\Vert_\spp \le 
C \Big\Vert \sqrt{G} \big\vert D_0(b) \big\vert^{-1} \Big\Vert_\spp + C \big\Vert \sqrt{G} T_1(\lambda) \big\Vert_\spp.
\end{equation} 

a) Firstly, we estimate the second term of the r.h.s. of \eqref{eq4,26}. Using \eqref{eq4,24}, we find that there exists a constant 
$C > 0$ such that
\begin{equation}\label{eq4,261}
\Big\Vert \sqrt{G} T_1(\lambda) \Big\Vert_\spp \le C \big( |\lambda| + |\lambda|^2 \big) 
\Big\Vert \sqrt{G} \big( D_0(b)^{2} - \lambda^{2} \big)^{-1} \Big\Vert_\spp.
\end{equation}
This together with the identity \eqref{Dir:opecar} implies that
\begin{equation}\label{eq4,262}
\begin{split}
\Big\Vert & \sqrt{G} T_1(\lambda) \Big\Vert_\spp \\
& \le C \big( |\lambda| + |\lambda|^2 \big) \Bigg( \Big\Vert \sqrt{G} \big( \cho(b) + 1 - \lambda^2 \big)^{-1} \Big\Vert_\spp + 
\Big\Vert \sqrt{G} \big( \cho(b) + 2b + 1 - \lambda^2 \big)^{-1} \Big\Vert_\spp \Bigg).
\end{split}
\end{equation}
We have
\begin{equation}\label{eq4,263}
\left\Vert \sqrt{G} \big( \cho(b) + 1 - \lambda^2 \big)^{-1} \right\Vert_\spp \leq \left\Vert \sqrt{G} \big( \cho(b) + 1 \big)^{-1} 
\right\Vert_\spp \left\Vert \big( \cho(b) + 1 \big) \big( \cho(b) + 1 - \lambda^2 \big)^{-1} 
\right\Vert.
\end{equation}
Since $\sigma \big( \cho(b) + 1 \big) = \cup_{q = 0}^\infty \lbrace \Lambda_q + 1 \rbrace$, then, the spectral 
mapping theorem implies that
\begin{equation}\label{eq4,264}
\Big\Vert \big( \cho(b) + 1 \big) \big( \cho(b) + 1 - \lambda^2 \big)^{-1} \Big\Vert 
\leq \textup{sup}_{\varrho \in \sigma(\cho(b) + 1)} \left\vert \frac{\varrho}{\varrho - \lambda^2} \right\vert 
\le \Bigg( 1 + \frac{\vert \lambda \vert^2} {{\rm dist} \big( \lambda^2,\cup_{q = 0}^\infty \lbrace \Lambda_q + 1 \rbrace \big)} 
\Bigg).
\end{equation}
Thus, reasoning as in the proof of Proposition \ref{l_estsch}, it can be shown by using \eqref{eq4,263}, the diamagnetic inequality,
the standard criterion \cite[Theorem 4.1]{sim} and the interpolation argument, that
\begin{equation}\label{eq4,265}
\left\Vert \sqrt{G} \big( \cho(b) + 1 - \lambda^2 \big)^{-1} \right\Vert_\spp \leq C(p,b) \big\Vert \sqrt{G} \big\Vert_{{\rm L}^p} 
\Bigg( 1 + \frac{\vert \lambda \vert^2} {{\rm dist} \big( \lambda^2,\cup_{q = 0}^\infty \lbrace \Lambda_q + 1 \rbrace \big)} \Bigg).
\end{equation}
Similarly, we have
\begin{equation}\label{eq4,266}
\begin{split}
\Big\Vert & \sqrt{G} \big( \cho(b) + 2b + 1 - \lambda^2 \big)^{-1} \Big\Vert_\spp \\
& \leq \left\Vert \sqrt{G} \big( \cho(b) + 2b + 1 \big)^{-1} \right\Vert_\spp 
\left\Vert \big( \cho(b) + 2b + 1 \big) \big( \cho(b) + 2b + 1 - \lambda^2 \big)^{-1} \right\Vert.
\end{split}
\end{equation}
Since $\sigma \big( \cho(b) + 2b + 1 \big) = \cup_{q = 0}^\infty \lbrace \Lambda_q + 2b + 1 \rbrace$, then, the spectral 
mapping theorem implies that
\begin{equation}\label{eq4,267}
\begin{split}
\Big\Vert & \big( \cho(b) + 2b + 1 \big) \big( \cho(b) + 2b + 1 - \lambda^2 \big)^{-1} \Big\Vert \\ 
& \leq \textup{sup}_{\varrho \in \sigma(\cho(b) + 2b + 1)} \left\vert \frac{\varrho}{\varrho - \lambda^2} \right\vert 
\le \Bigg( 1 + \frac{\vert \lambda \vert^2} {{\rm dist} \big( \lambda^2,\cup_{q = 0}^\infty \lbrace \Lambda_q + 2b + 1 \rbrace \big)} 
\Bigg).
\end{split}
\end{equation}
Thus, reasoning as in the proof of Proposition \ref{l_estsch}, it can be shown by using \eqref{eq4,266}, the diamagnetic inequality,
the standard criterion \cite[Theorem 4.1]{sim} and the interpolation argument, that
\begin{equation}\label{eq4,268}
\left\Vert \sqrt{G} \big( \cho(b) + 2b + 1 - \lambda^2 \big)^{-1} \right\Vert_\spp \leq C(p,b) \big\Vert \sqrt{G} \big\Vert_{{\rm L}^p} 
\Bigg( 1 + \frac{\vert \lambda \vert^2} {{\rm dist} \big( \lambda^2,\cup_{q = 0}^\infty \lbrace \Lambda_q + 2b + 1 \rbrace \big)} \Bigg).
\end{equation}
By putting together bounds \eqref{eq4,262}, \eqref{eq4,265} and \eqref{eq4,268}, we get 
\begin{equation}\label{eq4,269}
\Big\Vert \sqrt{G} T_1(\lambda) \Big\Vert_\spp \le C(p,b) \big\Vert \sqrt{G} \big\Vert_{{\rm L}^p} 
\big( |\lambda| + |\lambda|^2 \big) \big( 2 + C_1(\lambda) + C_2(\lambda) \big),
\end{equation}
where $C_1(\lambda)$ and $C_2(\lambda)$ are defined by \eqref{est_cdi}.

b) Now, we estimate the first term $\big\Vert \sqrt{G} \big\vert D_0(b) \big\vert^{-1} \big\Vert_\spp$ of the r.h.s. of 
\eqref{eq4,26}. Thanks to \eqref{Dir:opecar} and the identity 
$\big\vert D_0(b) \big\vert^{-\alpha} = \big( D_0(b)^2 \big)^{-\frac{\alpha}{2}}$, $\alpha > 0$, it follows that
\begin{equation}\label{eq4,270}
\Big\Vert \sqrt{G} \big\vert D_0(b) \big\vert^{-\alpha} \Big\Vert_\spp 
\le \Bigg( \Big\Vert \sqrt{G} \big( \cho(b) + 1 \big)^{-\frac{\alpha}{2}} \Big\Vert_\spp + 
\Big\Vert \sqrt{G} \big( \cho(b) + 2b + 1 \big)^{-\frac{\alpha}{2}} \Big\Vert_\spp \Bigg).
\end{equation}
Thus, as in the proof of a) above, the use of the diamagnetic inequality, the standard criterion \cite[Theorem 4.1]{sim} 
and the interpolation argument, allows to show that for $\alpha p > 2$, each term of the r.h.s. of \eqref{eq4,270} is bounded by
$C(p,b,\alpha) \big\Vert \sqrt{G} \big\Vert_{{\rm L}^p}$, where $C(p,b,\alpha) > 0$ is a constant depending only on 
$p$, $b$ and $\alpha$. In particular, for $\alpha = 1$, we obtain
\begin{equation}\label{eq4,271}
\Big\Vert \sqrt{G} \big\vert D_0(b) \big\vert^{-1} \Big\Vert_\spp \le C(p,b) \big\Vert \sqrt{G} \big\Vert_{{\rm L}^p}.
\end{equation}
This together with bounds \eqref{eq4,26} and \eqref{eq4,269} give the proposition.
\end{prof}

\section{The discrete eigenvalues as zeros of a holomorphic function}\label{s03}

For further use, let us recall some useful concepts by following \cite[Section 4]{goh}. Let $\mathscr{H}$ be a Hilbert space as above. We denote 
$\mathscr{L}(\mathscr{H})$ (resp. ${\rm GL}(\mathscr{H})$) the set of bounded (resp. invertible) operators in $\mathscr{H}$.

\begin{defi}
Let $\mathcal{U}$ be a neighbourhood of a fixed point $w \in \bc$, and 
$F : \mathcal{U} \setminus \lbrace w \rbrace \longrightarrow \mathscr{L}(\mathscr{H})$ 
be a holomorphic operator-valued function. The function $F$ is said to be finite 
meromorphic at $w$ if its Laurent expansion at $w$ has the form
$F(z) = \sum_{n = m}^{+\infty} (z - w)^n A_n$, $m > - \infty$,
where (if $m < 0$) the operators $A_m, \ldots, A_{-1}$ are of finite rank.
Moreover, if $A_0$ is a Fredholm operator, then, the function $F$ is said to be Fredholm 
at $w$. In that case, the Fredholm index of $A_0$ is called the Fredholm index of $F$ 
at $w$.
\end{defi}

\begin{prop}{\cite[Proposition 4.1.4]{goh}}\label{p,a1}
Let $\mathcal{D} \subseteq \mathbb{C}$ be a connected open set, $Z \subseteq \mathcal{D}$ 
be a closed and discrete subset of $\mathcal{D}$, and $F : \mathcal{D} \longrightarrow 
\mathscr{L}(\mathscr{H})$ be a holomorphic operator-valued function in $\mathcal{D} \backslash 
Z$. Assume that
$F$ is finite meromorphic on $\mathcal{D}$ (i.e. it is finite meromorphic near each 
point of $Z$),
$F$ is Fredholm at each point of $\mathcal{D}$,
and there exists $w_0 \in \mathcal{D} \backslash Z$ such that $F(w_0)$ is invertible. 
Then, there exists a closed and discrete subset $Z'$ of $\mathcal{D}$ such that
$Z \subseteq Z'$,
$F(z)$ is invertible for each $z \in \mathcal{D} \backslash Z'$,
$F^{-1} : \mathcal{D} \backslash Z' \longrightarrow {\rm GL}(\mathscr{H})$ is finite
meromorphic and Fredholm at each point of $\mathcal{D}$.
\end{prop}

In the setting of Proposition \ref{p,a1}, we define the characteristic values of $F$ and their multiplicities as follows:

\begin{defi}\label{d,a1}
The points of $Z'$ where the function $F$ or $F^{-1}$ is not holomorphic are called the
characteristic values of $F$. The multiplicity of a characteristic value $w_0$ is 
defined by
\begin{equation}
{\rm mult}(w_0) := \frac{1}{2i\pi} \textup{Tr} \int_{\vert w - w_0 \vert = \rho} 
F'(z)F(z)^{-1} dz,
\end{equation}
where $\rho > 0$ is chosen small enough so that $\big\lbrace w \in \bc : \vert w - 
w_0 \vert \leq \rho \big\rbrace \cap Z' = \lbrace w_0 \rbrace$.
\end{defi}

According to Definition \ref{d,a1}, if the function $F$ is holomorphic in $\mathcal{D}$,
then, the characteristic values of $F$ are just the complex numbers $w$ where the operator 
$F(w)$ is not invertible. Then, results of \cite{go} and \cite[Section 4]{goh} imply that 
${\rm mult}(w)$ is an integer.
Let $\Omega \subseteq \mathcal{D}$ be a connected domain with boundary $\partial \Omega$ 
not intersecting $Z'$. The sum of the multiplicities of the characteristic values of
the function $F$ lying in $\Omega$ is called {\it the index of $F$ with respect to the 
contour $\partial \Omega$} and is defined by 
\begin{equation}\label{eqa,2}
{\rm Ind}_{\partial \Omega} \, F := \frac{1}{2i\pi} \textup{Tr} \int_{\partial \Omega} 
F'(z)F(z)^{-1} dz = \frac{1}{2i\pi} \textup{Tr} \int_{\partial \Omega} F(z)^{-1} F'(z) dz.
\end{equation} 

In order to simplify the presentation and to shorten the article, we will treat simultaneously the three Hamiltonians. 
Hence, we recall that $\mathcal{H}_V(b)$ denotes the operators $\ch_V(b)$, $P_V(b)$ and $D_V(b)$. Thus, by \eqref{eq:spsch}, 
\eqref{eq:sppau} and \eqref{eq:spdi}, we have
\begin{equation}\label{eq:spcom}
\sigma \big( \mathcal{H}_0(b) \big) = \sigma_{\rm ess} \big( \mathcal{H}_0(b) \big) 
= \begin{cases} 
\cup_{q=0}^{\infty} \lbrace \Lambda_q \rbrace & \text{if } \: \: \mathcal{H}_0(b) 
= \cho(b) \: {\rm or} \: P_0(b), \\ 
\big\lbrace \cup_{q=1}^{\infty} \big\lbrace \Lambda_q^- \big\rbrace \big\rbrace 
\bigcup \big\lbrace \cup_{q=0}^{\infty} \big\lbrace \Lambda_q^+ \big\rbrace \big\rbrace 
& \text{if } \: \: \mathcal{H}_0(b) = D_0(b),
\end{cases}
\end{equation}
where $\Lambda_q = 2bq$ and $\Lambda_q^\pm = \pm \sqrt{2bq + 1}$ are the {\bf DLPLs}. In the sequel, w.r.t. \eqref{eq:spcom}, we will write 
$$
\sigma \big( \mathcal{H}_0(b) \big) = \sigma_{\rm ess} \big( \mathcal{H}_0(b) \big)
= \cup_{q=0}^{\infty} \big\lbrace \Lambda_q^\# \big\rbrace.
$$
${\bf P}_q^\#$, $q \ge 0$, will denote the orthogonal projection onto 
${\rm Ker} \, \big( \mathcal{H}_0(b) - \Lambda_q^\# \big)$, and ${\bf Q}_q^\#$, $q \ge 0$, will 
denote the orthogonal projection onto $\bigcup_{j \neq q} {\rm Ker} \, \big( \mathcal{H}_0(b) - \Lambda_j^\# \big)$. 
Thus, ${\bf Q}_q^\# = I - {\bf P}_q^\#$.

For a fixed spectral threshold $\Lambda_q^\# \in \cup_{q=0}^{\infty} 
\lbrace \Lambda_q \rbrace$, let
\begin{equation}
0 < \varepsilon < 2b.
\end{equation}
In the case $\Lambda_q^\# = \Lambda_q^\pm \in \big\lbrace \cup_{q=1}^{\infty} \big\lbrace \Lambda_q^- \big\rbrace \big\rbrace \bigcup \big\lbrace \cup_{q=0}^{\infty} \big\lbrace \Lambda_q^+ 
\big\rbrace \big\rbrace$ fixed, we impose that
\begin{equation}
0 < \varepsilon < \begin{cases} 
\sqrt{2b + 1} - 1 & \text{for } \: q = 0, \\ 
\min \big( \Lambda_q^\# - \Lambda_-,\Lambda_+ - 
\Lambda_q^\# \big) & \text{for } \: q \ge 1,
\end{cases}
\end{equation}
where $\Lambda_\pm$ denote the {\bf DLLs} respectively on the right and the left on $\Lambda_q^\#$. Hence, we define
$\mathscr{D}_{q} (\varepsilon)^{\ast} := \big\lbrace \lambda \in \mathbb{C} : 0 < \big\vert 
\Lambda_q^\# - \lambda \big\vert < \varepsilon \big\rbrace$.
Put the change of variables $\Lambda_q^\# - \lambda = k$ and introduce 
$
\mathscr{D} \big( 0,\varepsilon \big)^{\ast} := \big\lbrace k \in \mathbb{C} : 
0 < \vert k \vert < \varepsilon \big\rbrace$.
Thus, $\mathscr{D}_{q}(\varepsilon)^{\ast}$ can be parametrized by 
\begin{equation}\label{eq:lamb}
\lambda = \lambda_{q}(k) := \Lambda_{q}^\# - k, \quad k \in \mathscr{D} 
\big( 0,\varepsilon \big)^{\ast},
\end{equation}
and we have the relation
$\mathscr{D}_{q}(\varepsilon)^{\ast} = \Lambda_q^\# + \mathscr{D} \big( 0,\varepsilon \big)^{\ast}$.
We have the following proposition:

\begin{prop}\label{p3,1} 
Let $V = V_\omega$ satisfy the assumptions of Theorems \ref{t1,2}, \ref{t1,3} or \ref{t1,4}. Then, for any fixed 
spectral threshold $\Lambda_q^\#$, $q \ge 0$, the operator-valued function
$$
\mathscr{D}(0,\varepsilon)^{\ast} \ni k \longmapsto {\bf T}_{V_\omega} \big( \lambda_q(k) 
\big) := \pm \omega \sqrt{\vert W \vert} \big( \mathcal{H}_0(b) - \lambda_q(k) \big)^{-1} 
\sqrt{\vert W \vert}
$$ 
is analytic with values in the Schatten-von Neumann class $\spp$.
\end{prop} 

\noindent
\begin{prof}
Assume that $V = V_\omega$ satisfy the assumptions of Theorems \ref{t1,2}, \ref{t1,3} or \ref{t1,4}. Then, thanks to Propositions \ref{l_estsch}, 
\ref{l_estpau} and \ref{l_estdi}, together with 
$\lambda_q(k) \in \rho \big( \mathcal{H}_0(b) \big) = \bc \setminus \cup_{q=0}^{\infty}  \big\lbrace \Lambda_q^\# \big\rbrace$ for 
$k \in \mathscr{D}(0,\varepsilon)^{\ast}$, we have ${\bf T}_{V_\omega} \big( \lambda_q(k) \big) \in \spp$.

Let us show the analyticity of the map $\mathscr{D}(0,\varepsilon)^{\ast} \ni k \longmapsto 
{\bf T}_{V_\omega} \big( \lambda_q(k) \big)$. We have, using \eqref{eq:lamb},
\begin{align}
\sqrt{\vert W \vert} & \big( \mathcal{H}_0(b) - \lambda_q(k) \big)^{-1} \sqrt{\vert W \vert} 
\nonumber \\ 
& = \sqrt{\vert W \vert} {\bf P}_q^\# \big( \mathcal{H}_0(b) - \lambda_q(k) \big)^{-1} \sqrt{\vert W \vert} + 
\sqrt{\vert W \vert} {\bf Q}_q^\# \big( \mathcal{H}_0(b) - \lambda_q(k) \big)^{-1} \sqrt{\vert W \vert} \nonumber \\
& = k^{-1} \sqrt{\vert W \vert} {\bf P}_q^\# \sqrt{\vert W \vert} + \sqrt{\vert W \vert} {\bf Q}_q^\# 
\big( \mathcal{H}_0(b) - \lambda_q(k) \big)^{-1} \sqrt{\vert W \vert}.
\label{eq3,2}
\end{align}
Now, each term of the sum \eqref{eq3,2} is analytic in $\mathscr{D}(0,\varepsilon)^{\ast}$. 
Then, so is the map $\mathscr{D}(0,\varepsilon)^{\ast} \ni k \longmapsto {\bf T}_{V_\omega} 
\big( \lambda_q(k) \big)$. This concludes the proof.
\end{prof}

Propositions \ref{l_estsch}, \ref{l_estpau} and \ref{l_estdi} imply that the operator $\omega W \big( \mathcal{H}_0(b) - \lambda \big)^{-1}$ is of 
class $\spp$, $p \geq 2$, for $\lambda \in \rho \big( \mathcal{H}_0(b) \big)$. Consequently, we can introduce the $\lceil p \rceil$-regularized determinant 
\begin{equation}\label{eq3.3}
\begin{split}
& {\rm det}_{\lceil p \rceil} \Big( I + \omega W \big( \mathcal{H}_0(b) - \lambda 
\big)^{-1} \Big) \\
& := \det \left\lbrace \Big( I + \omega W \big( \mathcal{H}_0(b) - \lambda \big)^{-1} 
\Big) \exp \left( \sum_{k=1}^{\lceil p \rceil-1} \frac{\Big( - \omega W \big( \mathcal{H}_0(b) 
- \lambda \big)^{-1} \Big)^{k}}{k} \right) \right\rbrace,
\end{split}
\end{equation}
where $\lceil p \rceil := \min \big\lbrace n \in \mathbb{N} : n \geq p \big\rbrace$. It 
is well known, see for instance \cite[Chap. 9]{sim}, that we have the characterization 
\begin{equation}\label{eq3.4}
\lambda \in \sigma_{\rm disc} \big( \mathcal{H}_{V_\omega}(b) \big) \Leftrightarrow 
f_{p}(\lambda) := \textup{det}_{\lceil p \rceil} \Big( I + \omega W \big( \mathcal{H}_0(b) 
- \lambda \big)^{-1} \Big) = 0.
\end{equation}
Moreover, if the operator $\omega W \big( \mathcal{H}_0(b) - \lambda \big)^{-1}$ is holomorphic in a domain $\Omega$, then so is the function $f_{p}(\lambda)$ 
in $\Omega$, and the algebraic multiplicity of $\lambda \in \sigma_{\rm disc} \big( \mathcal{H}_{V_\omega}(b) \big)$ is equal to its order as zero of the 
regularized determinant $f_{p}(\lambda)$.


\begin{prop}\label{p3,2} 
Let $V = V_\omega$ satisfy the assumptions of Theorems \ref{t1,2}, \ref{t1,3} or \ref{t1,4}. Let ${\bf T}_{V_\omega} \big( \lambda_q(k) \big)$ be 
the operator defined in Proposition \ref{p3,1}. Then, for $k_{0} \in \mathscr{D}(0,\varepsilon)^{\ast}$, the following assertions are equivalent:
\begin{itemize}
\item[(i)] $\lambda_q(k_{0}) = \Lambda_q^\# - k_{0} \in \mathscr{D}_{q}(\varepsilon)^{\ast}$ is a discrete eigenvalue
of $\mathcal{H}_{V_\omega}(b)$,

\item[(ii)] $\textup{det}_{\lceil p \rceil} \Big( I + {\bf T}_{V_\omega} 
\big( \lambda_q (k_{0}) \big) \Big) = 0$,

\item[(iii)] $-1$ is an eigenvalue of ${\bf T}_{V_\omega} \big( \lambda_q(k_{0}) \big)$.
Moreover, the following equality happens
\begin{equation}\label{eq3,5}
\textup{mult} \big( \lambda_q(k_{0}) \big) = {\rm Ind}_{\gamma} \hspace{0.5mm} \Big( I + 
{\bf T}_{V_\omega}\big( \lambda_q(\cdot) \big) \Big),
\end{equation}
where $\gamma$ is a small contour positively oriented containing $k_{0}$ as the unique point 
$k$ satisfying $\lambda_q(k)$ is a discrete eigenvalue of $\mathcal{H}_{V_\omega}(b)$.
\end{itemize}
\end{prop}

\noindent
\begin{prof}
(i) $\Leftrightarrow$ (ii) follows from \eqref{eq3.4} and the equality 
$$
\textup{det}_{\lceil p \rceil} \Big( I + \omega W \big( \mathcal{H}_0(b) - \lambda \big)^{-1} 
\Big) = \textup{det}_{\lceil p \rceil} \Big( I \pm \omega \sqrt{\vert W \vert} 
\big( \mathcal{H}_0(b) - \lambda \big)^{-1} \sqrt{\vert W \vert} \Big).
$$
(ii) $\Leftrightarrow$ (iii) is a direct consequence of the definition of 
$\textup{det}_{\lceil p \rceil} ( I + K)$, $K \in \spp$, similarly to \eqref{eq3.3}.

Let us prove \eqref{eq3,5}. Let $f_{p}(\lambda)$ be the function defined by \eqref{eq3.4}. By the discussion just after \eqref{eq3.4}, if $\gamma'$ is 
a small contour positively oriented containing $\lambda_q(k_{0})$ as the unique discrete eigenvalue of $\mathcal{H}_{V_\omega}(b)$, then, we have
\begin{equation}\label{eq3,6}
\textup{mult} \big( \lambda_q(k_{0}) \big) = {\rm ind}_{\gamma'} f_{p} = 
\frac{1}{2i\pi} \int_{\gamma'} \frac{f'(\lambda)}{f(\lambda)} d\lambda.
\end{equation}
Now, \eqref{eq3,5} follows from the equality 
$
{\rm ind}_{\gamma'} f_{p} = {\rm Ind}_{\gamma} \, \big( I + {\bf T}_{V_\omega} 
\big( \lambda_q(\cdot) \big) \big)
$,
see for instance the identity (2.6) of \cite{bo} for more details.
\end{prof}

\section{Proof of Theorems \ref{t1,2}, \ref{t1,3} and \ref{t1,4}}\label{s4}

We conserve the notations introduced in the previous Section. By \eqref{eq3,2}, for 
$\lambda_q(k) \in \mathscr{D}_q(\varepsilon)^\ast$, $k \in \mathscr{D}(0,\varepsilon)^\ast$, we have
\begin{equation}\label{eq4,1}
{\bf T}_{V_\omega} \big( \lambda_q(k) \big) = \pm \omega \frac{\sqrt{\vert W \vert} {\bf P}_q^\# 
\sqrt{\vert W \vert}}{k} \pm \omega \sqrt{\vert W \vert} {\bf Q}_q^\# \big( \mathcal{H}_0(b) - 
\lambda_q(k) \big)^{-1} \sqrt{\vert W \vert}.
\end{equation}
Thus, the following proposition holds:

\begin{prop}\label{p4,1} 
Let $V = V_\omega$ satisfy the assumptions of Theorems \ref{t1,2}, \ref{t1,3} or \ref{t1,4}. Let ${\bf T}_{V_\omega} \big( \lambda_q(k) \big)$ be 
the operator defined in Proposition \ref{p3,1}. Then, we have
\begin{equation}\label{eq4,2}
{\bf T}_{V_\omega} \big( \lambda_q(k) \big) = \pm \omega \frac{\sqrt{\vert W \vert} {\bf P}_q^\# 
\sqrt{\vert W \vert}}{k} \pm \omega {\bf A}_q(k),
\end{equation}
where the operator
${\bf A}_q(k) := \sqrt{\vert W \vert} {\bf Q}_q^\# \big( \mathcal{H}_0(b) - \lambda_q(k) \big)^{-1} 
\sqrt{\vert W \vert} \in \sinf$
is holomorphic in $\mathscr{D}(0,\varepsilon) := \mathscr{D}(0,\varepsilon)^{\ast} \cup \lbrace 0 \rbrace$.
\end{prop} 

Now, we formulate Proposition \ref{p3,2} in terms of characteristic values, see Definition \ref{d,a1}.

\begin{prop}\label{p5,1} 
Let $V = V_\omega$ satisfy the assumptions of Theorems \ref{t1,2}, \ref{t1,3} or \ref{t1,4}. Then, for $k_{0} \in \mathscr{D}(0,\varepsilon)^{\ast}$, 
the following assertions are equivalent:
\begin{itemize}
\item[(i)] $\lambda_q(k_{0}) = \Lambda_q^\# - k_{0} \in \mathscr{D}_{q}(\varepsilon)^{\ast}$ is a discrete eigenvalue of $\mathcal{H}_{V_\omega}(b)$,

\item[(ii)] $k_{0}$ is a characteristic value of $I + {\bf T}_{V_\omega} \big( \lambda_q(k) \big)$. Moreover, 
we have $\textup{mult} \big( \lambda_q(k_{0}) \big) = \textup{mult} (k_{0})$.
\end{itemize}
\end{prop}

By setting 
\begin{equation}\label{eq1,9}
{\bf \widetilde A}_q(k) := \sqrt{\vert W \vert} {\bf P}_q^\# \sqrt{\vert W \vert} + k{\bf A}_q(k),
\end{equation}
it follows from Proposition \ref{p5,1} that the study of the discrete eigenvalues $\lambda_q(k)$ near a 
fixed spectral threshold $\Lambda_q^\#$, $q \ge 0$, can be reduced to that of the characteristic values of 
\begin{equation}\label{eqcv}
I + {\bf T}_{V_\omega} \big( \lambda_q(k) \big) = I \pm \omega \frac{{\bf \widetilde A}_q(k)}{k}
= I - \frac{{\bf \widetilde A}_q^{(\omega)}(z)}{z},
\end{equation}
where $z = \mp k/\omega$ and ${\bf \widetilde A}_q^{(\omega)}(z) := {\bf \widetilde A}_q (\mp \omega z)$. In 
particular, we have ${\bf \widetilde A}_q^{(\omega)}(0) = {\bf \widetilde A}_q(0) = \sqrt{\vert W \vert} 
{\bf P}_q^\# \sqrt{\vert W \vert}$. Furthermore, we have 
$
\big( {\bf \widetilde A}_q^{(\omega)} \big)'(z) = \mp \omega {\bf \widetilde A}_q'(\mp \omega z)
$
implying that $\big( {\bf \widetilde A}_q^{(\omega)} \big)'(0) = \mp \omega {\bf \widetilde A}_q'(0)$. Let 
$\widetilde {\bf \Pi}_q$ denote the orthogonal projection onto $\text{Ker} \, {\bf \widetilde A}_q(0)$, and note that 
${\bf \widetilde A}_q'(0) \widetilde {\bf \Pi}_q$ is a compact operator. Thus, there exists a discrete set 
$$
\bc^\ast \supset \widetilde \Sigma_q := 
\begin{cases} 
\Sigma_q & \text{if } \: {\mathcal{H}_V(b) = \ch_V(b)}, \\ 
\Xi_q & \text{if } \: {\mathcal{H}_V(b) = P_V(b)}, \\
\Sigma_q^\pm & \text{if } \: {\mathcal{H}_V(b) = D_V(b)},
\end{cases}
$$
such that the operator $I - \big( {\bf \widetilde A}_q^{(\omega)} \big)'(0) \widetilde {\bf \Pi}_q = I \pm \omega 
{\bf \widetilde A}_q'(0) \widetilde {\bf \Pi}_q$ is invertible for each $\omega \in \bc^\ast \setminus \widetilde 
\Sigma_q$.

Thus, (i) of Theorems \ref{t1,2}, \ref{t1,3} and \ref{t1,4} is an immediate consequence of \cite[Corollary 3.4. 
(i) and (ii)]{bo} with $z = \mp k/\omega$. More precisely, the discrete eigenvalues $\lambda_q(k)$ satisfy
\begin{equation}\label{eq5,20}
\mp \Re \left( \frac{k}{\omega} \right) \geq 0, \qquad k \in \mp \omega 
\overline{\mathcal{S}(\delta,r,r_{0})},
\end{equation}
for any $\delta > 0$, with the sector $\mathcal{S}(\delta,r,r_{0})$ defined by \eqref{eq1,10}.

Now, Proposition \ref{p5,1} together with \eqref{eqcv} show that $\lambda_q(k)$ is a discrete eigenvalue of $\mathcal{H}_{V_\omega}(b)$ if 
and only if $z = \mp k/\omega$ is a characteristic value of ${\bf \widetilde A}_q^{(\omega)}(z) = {\bf \widetilde A}_q (\mp \omega z)$, with the 
same multiplicity. In the sequel, we denote this set characteristic values by ${\rm Char}(\bullet)$. Futhermore, \eqref{eq5,20} shows 
that for $\vert \omega \vert r < \vert k \vert < \vert \omega \vert r_0$, the characteristic values 
$z = \mp k/\omega$ are concentrated in a sector $\mathcal{S}(\delta,r,r_{0})$ for any $\delta > 0$. In 
particular, for $r \searrow 0$, we have
\begin{equation}\label{eq5,3}
\# \Big\lbrace \lambda_{q}(k) \in \sigma_{\textup{disc}} \big( \mathcal{H}_{V}(b) \big) : \vert \omega \vert r < \vert k \vert < \vert \omega \vert r_{0} \Big\rbrace 
= \# \Big\lbrace z = \mp k/\omega \in {\rm Char}(\bullet) \cap \mathcal{S}(\delta,r,r_{0}) \Big\rbrace 
+ \mathcal{O}(1).
\end{equation}
Due to \eqref{eq:tosch}, \eqref{eq:topau} and \eqref{eq:todi}, we have ${\bf \widetilde A}_q^{(\omega)}(0) = 
\sqrt{\vert W \vert} {\bf P}_q^\# \sqrt{\vert W \vert} \in \spp$. Then, if 
${\rank} \big( {\bf \widetilde A}_q^{(\omega)}(0) \big) = \infty$, \cite[Corollary 3.9]{bo} implies that 
there exists a sequence $(\eta_\ell)_\ell$ of positive number tending to zero such that
\begin{equation}\label{eq5,4}
\begin{split}
\# \Big\lbrace z = \mp k/\omega \in {\rm Char}(\bullet) \cap \mathcal{S}(\delta,\eta_\ell,r_{0}) \Big\rbrace 
& = \textup{Tr} \, \mathbf{1}_{[\eta_\ell,\infty)} \left( \sqrt{\vert W \vert} {\bf P}_q^\# \sqrt{\vert W \vert} \right) \big( 1 + o(1) \big) \\
& = \textup{Tr} \, \mathbf{1}_{[\eta_\ell,\infty)} \left( {\bf P}_q^\# \vert W \vert {\bf P}_q^\# \right) 
\big( 1 + o(1) \big), \quad \ell \longrightarrow \infty.
\end{split}
\end{equation}
Thus, by putting together \eqref{eq5,3} and \eqref{eq5,4}, it follows Theorem \ref{t1,2} (ii), Theorem \ref{t1,3} 
(ii)-(iii), and Theorem \ref{t1,4} (ii), with
$$
\eta_\ell := 
\begin{cases} 
r_\ell & \text{if } \: {\mathcal{H}_V(b) = \ch_V(b)}, \\ 
\mu_\ell \, \: or \, \: \nu_\ell & \text{if } \: {\mathcal{H}_V(b) = P_V(b)}, \\
\gamma_\ell & \text{if } \: {\mathcal{H}_V(b) = D_V(b)}.
\end{cases}
$$


\begin{thebibliography}{99}

\bibitem[AHS78]{avr}
   \textsc{J. Avron, I. Herbst, B. Simon},
    \textit{Schr\"odinger operators with magnetic fields. I. General interactions},
    Duke Math. J. \textbf{45} (1978), 847-883.

\bibitem[AGH]{alm}
    \textsc{J. Almog, D. S. Grebenkov, B. Helffer},
    \textit{Spectral semi-classical analysis of a complex Schr\"odinger operator in exterior domains},
    arXiv: 1708.02926.
    
\bibitem[BBR14]{bo}
    \textsc{J.-F. Bony, V. Bruneau, G. Raikov},
    \textit{Counting function of characteristic values and magnetic resonances},
    Commun. PDE. \textbf{39} (2014), 274-305. 
        
\bibitem[BGK09]{bor}
    \textsc{A. Borichev, L. Golinskii, S. Kupin},
    \textit{A Blaschke-type condition and its application to complex Jacobi matrices},
    Bull. London Math. Soc. \textbf{41} (2009), 117-123.

\bibitem[B\"og17]{bog}
    \textsc{S. B\"ogli},
    \textit{Schr\"odinger operators with non-zero accumulation points of complex eigenvalues},
    Comm. Math. Phys. {\bf 352} (2017), no. 2, 629-639.    
            

\bibitem[CLT14]{cue}
    \textsc{J.-C Cuenin, A. Laptev, C. Tretter},
    \textit{Eigenvalues estimates for non-selfadjoint Dirac operators on the real line},
    Ann. Henri Poincaré, \textbf{15}(4), (2014), 707-736.
        
               


\bibitem[DR01]{dr}
    \textsc{M. Dimassi, G. Raikov},
    \textit{Spectral asymptotics for quantum Hamiltonians in strong magnetic fields},
    Cubo Matem\'atica Educacional, \textbf{3} (2001), 317-391. 
            
\bibitem[ET]{eng}
    \textsc{C. Engstr\"om, A. Torshage},
    \textit{Accumulation of complex eigenvalues of a class of analytic operator functions},
    arXiv: 1709.01462    


\bibitem[Fol84]{fol}
    \textsc{G. B. Folland},
    \textit{Real analysis Modern techniques and their applications},
    Pure and Apllied Mathematics, (1984), John Whiley and Sons. 
    
\bibitem[GS71]{go}
     \textsc{I. Gohberg, E. I. Sigal},
     \textit{An operator generalization of the logarithmic residue theorem and Rouch\'e's theorem},
     Mat. Sb. (N.S.) \textbf{84} (126) (1971), 607-629. 
     
\bibitem[GGK90]{goh}
    \textsc{I. Gohberg, S. Goldberg, M. A. Kaashoek},
    \textit{Classes of Linear Operators},
    Operator Theory, Advances and Applications, vol. \textbf{49} Birkh\"auser Verlag, 1990.

\bibitem[GL09]{gole}
     \textsc{I. Gohberg, J. Leiterer},
     \textit{Holomorphic operator  functions of one variable  and applications},
     Operator Theory, Advances and Applications, vol. \textbf{192} Birkh\"auser Verlag, 2009, Methods
     from complex analysis in several variables.
         
\bibitem[GGK00]{gohb}
    \textsc{I. Gohberg, S. Goldberg, N. Krupnik},
    \textit{Traces and Determinants of Linear Operators},
    Operator Theory, Advances and Applications, vol. \textbf{116} Birkh\"auser Verlag, 2000.
    
        
    


\bibitem[LT75]{lt1}
    \textsc{E. H. Lieb, W. Thirring},
    \textit{Bound for the kinetic energy of fermions which proves the stability of matter},
    Phys. Rev. lett. \textbf{35} (1975), 687-689. Errata \textbf{35} (1975), 1116.  

\bibitem[Lun09]{lun}
   \textsc{A. Lunardi},
   \textit{Interpolation Theory},
   Appunti Lecture Notes, \textbf{9} 2009, Edizioni Della Normale.
   
\bibitem[MR03]{mel}
    \textsc{M. Melgaard, G. Rozenblum}
    \textit{Eigenvalue asymptotics for weakly perturbed Dirac and Schr\"odinger operators 
    with constant magnetic fields of full rank},    
    Commun. PDE. \textbf{28} (2003), 697-736.
    
\bibitem[Pav67]{pav}
    \textsc{B. S. Pavlov},
    \textit{On a non-selfadjoint Schr\"odinger operator. II}, (Russian),
    Problems of Mathematical Physics, Izdat. Leningrad. Univ. {\bf 2} (1967), 133-157.

\bibitem[Rai90]{ra}
    \textsc{G. D. Raikov},
    \textit{Eigenvalue asymptotics for the Schr\"odinger operator with homogeneous magnetic 
    potential and decreasing electric potential. I. Behaviour near the essential spectrum tips}, 
    Commun. PDE. \textbf{15} (1990), 407-434.  
 
\bibitem[RW02]{raik}    
    \textsc{G. D. Raikov, S. Warzel}, 
    \textit{Quasi-classical versus non-classical spectral asymptotics for magnetic Schr\"odinger 
    operators with decreasing electric potentials},
     Rev. in Math. Physics, \textbf{14}(10) (2002) 1051-1072. 
         
\bibitem[RS79]{ree}
    \textsc{M. Reed, B. Simon},
    \textit{Scattering Theory III},
    Methods of Modern Mathematical Physics, (1979), Academic Press, INC.  

\bibitem[Rie26]{rie}  
   \textsc{M. Riesz}, 
   \textit{Sur les maxima des formes bilin\'eaires et sur les fonctionnelles lin\'eaires}, 
   Acta Math. \textbf{49} (1926), 465-497.
   
    

\bibitem[Sam17]{dio1-17}
    \textsc{D. Sambou},
    \textit{On eigenvalue accumulation for non-self-adjoint magnetic operators}, 
    J. Maths Pures et Appl. \textbf{108} (2017), 306-332. 

\bibitem[Sa17]{dio2-17}
    \textsc{D. Sambou},
    \textit{A simple criterion for the existence of nonreal eigenvalues for a class of 2D and 3D Pauli operators},
    Linear Alg. and its Appli. \textbf{529} (12) (2017), 51-88.   
    
\bibitem[Shi91]{shi}
    \textsc{I. Shigekawa},
    \textit{Spectral analysis of Schr\"{o}dinger operators with magnetic fields for a spin $\frac{1}{2}$ particule},
    J. Funct. Anal. \textbf{101} (1991), 255-285.
     
\bibitem[Sim79]{sim}
    \textsc{B. Simon},
    \textit{Trace ideals and their applications},
    Lond. Math. Soc. Lect. Not. Series, \textbf{35} (1979), Cambridge University Press.  

\bibitem[Tho39]{tho}  
    \textsc{G. O. Thorin}, 
    \textit{An extension of a convexity theorem due to M. Riesz}, 
    Kungl. Fysiografiska Saellskapet i Lund Forhaendlinger \textbf{8} (1939), no. 14. 
    
\bibitem[Wan11]{wan}
    \textsc{X. P. Wang},
    \textit{Number of eigenvalues for a class of non-selfadjoint Schrödinger operators},
    J. Maths Pures et Appl. \textbf{96}(9) (2011), no. 5, 409-422.  
   
   
    
\end{thebibliography}
\end{document}